\documentclass{aa}
\usepackage[dvips]{graphicx}
\usepackage{txfonts}

\def\kms{\mbox{km~s$^{-1}$}}

\def\Vlsr{$V_{\rm LSR}$}

\def\jykms{~Jy~beam$^{-1}$\,km~s$^{-1}$}

\def\jdo{\mbox{$J$=12$\rightarrow$11}}

\def\juz{\mbox{$J$=1$\rightarrow$0}}

\def\iras{IRAS~20050+2720}

\begin{document}

\title{On the nature of outflows in intermediate-mass protostars: a case study
of IRAS~20050+2720}

\author{M.\ T.\ Beltr\'an \inst{1} \and R.\ Estalella \inst{1} \and J.\ M.\ Girart \inst{2} \and P.\ T.\
P.\ Ho \inst{3, 4} \and G.\ Anglada \inst{5}} 

\offprints{M. T. Beltr\'an, \email{mbeltran@am.ub.es}}

\institute{
Departament d'Astronomia i Meteorologia, Universitat de Barcelona, Mart\'{\i} i Franqu\`es
1, 08028 Barcelona, Catalunya, Spain  
\and
Institut de Ci\`encies de l'Espai (CSIC-IEEC), Campus UAB, Facultat de Ci\`encies, Torre C-5, 08193, Bellaterra,
Catalunya, Spain
\and
Harvard-Smithsonian Center for Astrophysics, 60 Garden Street, Cambridge, MA 02138, USA
\and 
Academia Sinica, Institute of Astronomy and Astrophysics, P.O.\ Box 23--141, Taipei 106, Taiwan
\and
Instituto de Astrof\'{\i}sica de Andaluc\'{\i}a, CSIC, Camino Bajo de Hu\'etor 50, 18008 Granada, Spain}

\date{Received date; accepted date}

\titlerunning{On the nature of outflows in intermediate-mass protostars}
\authorrunning{Beltr\'an et al.}

\abstract
{This is the third of a series of papers devoted to study in detail and with high-angular
resolution intermediate-mass molecular
outflows and their powering sources.}
{The aim of this paper is to study the intermediate-mass YSO \iras\ and its molecular outflow, and put
the results of this and the previous studied sources in the context of intermediate-mass star formation. }
{We carried out VLA observations of the 7~mm continuum emission, and OVRO observations of the 2.7~mm
continuum emission, CO (\juz), C$^{18}$O (\juz), and HC$_3$N~(\jdo) to map the core towards \iras. The
high-angular resolution of the observations allowed us  to derive the properties of the dust emission, the
molecular outflow, and the dense protostellar envelope. By adding this source to the sample of
intermediate-mass protostars with outflows, we compare their properties
and evolution with those of lower mass counterparts.}
{The 2.7~mm continuum emission has been resolved into three sources, labeled OVRO~1, OVRO~2, and OVRO~3.
Two of them, OVRO~1 and OVRO~2, have also been detected at 7~mm. OVRO~3, which is located close to the
C$^{18}$O emission peak, could be associated with \iras.  The mass of the sources, estimated from the
dust continuum  emission, is 6.5~$M_{\odot}$ for OVRO~1, 1.8~$M_{\odot}$ for OVRO~2, and 1.3~$M_{\odot}$
for OVRO~3. The CO~(\juz) emission traces two bipolar outflows within the OVRO field of
view, a roughly east-west bipolar
outflow, labeled A, driven by the intermediate-mass source OVRO~1, and a northeast-southwest bipolar
outflow, labeled B, probably powered
by a YSO engulfed in the circumstellar envelope surrounding OVRO~1.}
{The multiplicity of sources observed towards \iras\ appears to be typical of intermediate-mass
protostars, which form in dense clustered environments. In some cases, as for example \iras,
intermediate-mass protostars would start forming after a first generation of low-mass stars has completed
their main accretion phase. Regarding the properties of intermediate-mass protostars and their outflows,
they are not  significantly different from those of low-mass stars. Although intermediate-mass outflows are
intrinsically more energetic than those driven by low-mass YSOs, when observed with high-angular resolution
they do not show intrinsically more complex morphologies. Intermediate-mass protostars do not form a
homogeneous group. Some objects are likely in an earlier evolutionary stage as suggested by the infrared
emission and the outflow properties.} 

\maketitle

\section{Introduction}

Molecular outflow is an ubiquitous phenomenon during the earliest stages of the formation of
stars of all masses and luminosities. During the last decades, many efforts have been devoted
to the study and description of the physical properties of embedded low-mass protostars and
their molecular outflows (e.g., Bachiller~\cite{bachiller96}; Richer et al.~\cite{richer00}).
In recent years,  high-mass molecular outflows have also been studied in detail, and many
surveys have been carried out towards massive star-forming regions to achieve a more accurate
picture of their morphology and properties (e.g., Shepherd \& Churchwell~\cite{shepherd96a},
\cite{shepherd96b}; Zhang et al.~\cite{zhang01}; Shepherd~\cite{shepherd05}).  In recent
years, there has been a growing interest for the Young Stellar Objects (YSOs) located at the
lower and the higher end of the stellar mass spectrum. Less attention has been paid to
intermediate-mass protostars, with  masses in the range $2~M_{\odot}\leq
M_\star\leq10~M_{\odot}$, which can power energetic molecular outflows. In fact, only few  deeply
embedded intermediate-mass protostars are known to date, and only a very limited number of
their outflows have been studied at high-angular resolution.

\begin{table*}
\caption[] {Parameters of the observations}
\label{tobs}
\begin{tabular}{lcccrcc}
\hline
&&&\multicolumn{2}{c}{Synthesized beam} \\
\cline{4-5}
&&\multicolumn{1}{c}{Frequency} &
\multicolumn{1}{c}{{\it HPBW}} &
\multicolumn{1}{c}{PA} &
\multicolumn{1}{c}{Spectral resolution}& 
\multicolumn{1}{c}{rms noise$^a$}\\
\multicolumn{1}{c}{Observation} &
\multicolumn{1}{c}{Telescope} &
\multicolumn{1}{c}{(GHz)} &
\multicolumn{1}{c}{(arcsec)} &
\multicolumn{1}{c}{(deg)} &
\multicolumn{1}{c}{(\kms)}& 
\multicolumn{1}{c}{(mJy~beam$^{-1}$)}
\\
\hline
2.7~mm continuum &OVRO &111.02 &$7.0\times6.4$ &$-49$ &$-$ &1.5 \\
CO (\juz) &OVRO  &115.27 &$6.8\times4.1$ &$-59$  &0.65 &55\\
C$^{18}$O (\juz) &OVRO &109.78  &$8.1\times6.0$ &$-43$  &0.68 &40\\
HC$_3$N (\jdo) &OVRO  &109.17 &$8.1\times6.0$ &$-47$  &0.69 &40\\
7~mm continuum &VLA &\phantom{1}43.34 &$4.9\times3.8$ &$-9$ &$-$ &0.16 \\
\hline 

\end{tabular}

(a) For the molecular line observations the 1~$\sigma$ noise is per channel.

\end{table*}

In order to study in detail intermediate-mass protostars and the outflows that they are powering and to
compare their morphology and evolution with those of low-mass stars, we have started a project to carry out
interferometric observations of dust and gas towards intermediate-mass star forming regions. The first two
regions observed were IRAS~21391+5802 (Beltr\'an et al.~\cite{beltran02}; hereafter Paper~I) and L1206
(Beltr\'an et al.~\cite{beltran06b}; hereafter Paper~II). In this third paper, we present a thorough
interferometric study  of the intermediate-mass protostar \iras\ and of its molecular outflow. \iras\ is a
$280~L_{\odot}$ (Froebrich~\cite{froebrich05}) YSO located in the Cygnus rift at a distance of 700~pc (Dame
\& Thaddeus~\cite{dame85}).Distances in the Cygnus region are highly uncertain for stellar objects
located at Galactic longitudes close to 90\degr, as the kinematic distances are not reliable (e.g.\
Schneider et al.~\cite{schneider06}). 
However, \iras\ has a Galactic longitude of $\sim$66\degr, being located
in a region where the Cygnus rift can be distinguished from the Cygnus X complex. Therefore, the kinematic distance
determined by  Dame \& Thaddeus~(\cite{dame85}) from CO line emission is quite reliable. \iras\ is
surrounded by a cluster of infrared stars (Chen et al.~\cite{chen97}). Bachiller et
al.~(\cite{bachiller95}) mapped a high-velocity multipolar CO outflow in the region, previously detected by
Wilking et al.~(\cite{wilking89}). \iras\ is deeply embedded in a core that has been observed in the
continuum at millimeter (Wilking et al.~\cite{wilking89}; Chen et al.~\cite{chen97}; Choi et
al.~\cite{choi99}; Chini et al.~\cite{chini01}; Furuya et al.~\cite{furuya05}; Beltr\'an et
al.~\cite{beltran06a}) and centimeter (Wilking et al.~\cite{wilking89}; Anglada et al.~\cite{anglada98a})
wavelengths, and also in different high-density tracers, such as HCO$^+$ and HCN (Choi et
al.~\cite{choi99}), NH$_3$ (Molinari et al. \cite{molinari96}), CS and N$_2$H$^+$ (Bachiller et
al.~\cite{bachiller95}; Williams \& Myers~\cite{williams99}), and $^{13}$CO and C$^{18}$O (Ridge et
al.~\cite{ridge03}). The source has been classified as an intermediate-mass Class 0/I protostar (Gregersen et
al.~\cite{gregersen97}; Chini et al.~\cite{chini01}; Froebrich~\cite{froebrich05}).

As shown by these previous studies, intermediate-mass outflows appear more collimated and less
complex than previously thought (e.g.\ Fuente et al.~\cite{fuente01}) when observed with high-angular
resolution. This argues for the need of high-angular observations to better understand
intermediate-mass protostars and their outflows. To accomplish the ultimate goal of our research
project, which is to study and compare the properties and evolution of intermediate-mass protostars
with those of their lower mass counterparts, in this last paper of the series, we put the results on
the three intermediate-mass protostars studied in the context of intermediate-mass star formation. 
To do this, we have compiled information on outflow properties that  have been derived from
interferometric observations to better compare them with the ones derived for the sources in our
study.

\section{Observations}
\label{obs}

\subsection{OVRO observations}

Millimeter interferometric observations of \iras\ at 2.7~mm were carried out with the
Owens Valley Radio Observatory (OVRO) Millimeter Array of six 10.4~m telescopes in the
L (Low) and  C (Compact) configurations, on April 30 and May~30, 2003, respectively. The
data taken in both array configurations were combined, resulting in baselines ranging
from 5.6 to 42.6\,$k\lambda$, which provided sensitivity to spatial structures from about
4$\farcs$8 to 37$''$. The digital correlator was configured to simultaneously observe 
the continuum emission and some molecular lines. Details of the observations are given
in Table~\ref{tobs}. The phase center was located at  $\alpha$(J2000) = $20^{\rm h}  
07^{\rm h} 5\fs80$,  $\delta$(J2000) = $+27\degr 28' 58\farcs1$. Bandpass calibration
was achieved by observing the quasars 3C84, 3C273, and 3C345. Amplitude and phase were
calibrated by observations of the nearby quasar J2015+71, whose flux density was
determined relative to Uranus. The uncertainty in the amplitude calibration is
estimated to be $\sim$$20\%$. The OVRO primary beam is $\sim$$64''$ (FWHM) at 115.27
GHz. The data were calibrated using the MMA package  developed for OVRO (Scoville et
al.~\cite{scoville93}). Reduction and analysis of the data were carried out using
standard procedures in the MIRIAD, AIPS, and GILDAS software packages.  We subtracted the
continuum from the line emission directly in the ({\sl u, v}) domain for C$^{18}$O
(\juz) and HC$_3$N (\jdo), and  in the image domain for CO (\juz).

\subsection{VLA observations}

The interferometric observations at 7~mm were carried out on January 5, 2002 using the Very
Large Array (VLA) of the National Radio Astronomy Observatory (NRAO)\footnote{NRAO is a
facility of the National Science Foundation operated under cooperative agreement by Associated
Universities, Inc.} in the D configuration. The baselines ranged
from 5.4 to 140\,$k\lambda$, which provided sensitivity to spatial structures from about
1$\farcs$5 to 37$''$.
The phase center was set to the position
$\alpha$(J2000) = $20^{\rm h}   07^{\rm h} 5\fs91$,  $\delta$(J2000) = $+27\degr 28'
58\farcs2$. Absolute flux calibration was achieved by observing 3C286, with an adopted flux
density of 1.45 Jy. Amplitude and phase were calibrated by observing 2023+318, which has a bootstrapped flux
of $0.48\pm0.01$~Jy. CLEANed maps were made using the task IMAGR of the AIPS software of the
NRAO, with the ROBUST parameter of Briggs~(\cite{briggs95}) set equal to 5, which is close to
natural weighting. We used a Gaussian taper in the ({\sl u, v}) domain when making the
maps in order to enhance the extended emission in the region. The resulting synthesized beam is $4\farcs9\times3\farcs8$ at PA =$-9\degr$,
and the rms noise is 0.16~mJy~beam$^{-1}$.

\begin{figure*}
\centerline{\includegraphics[angle=-90,width=14cm]{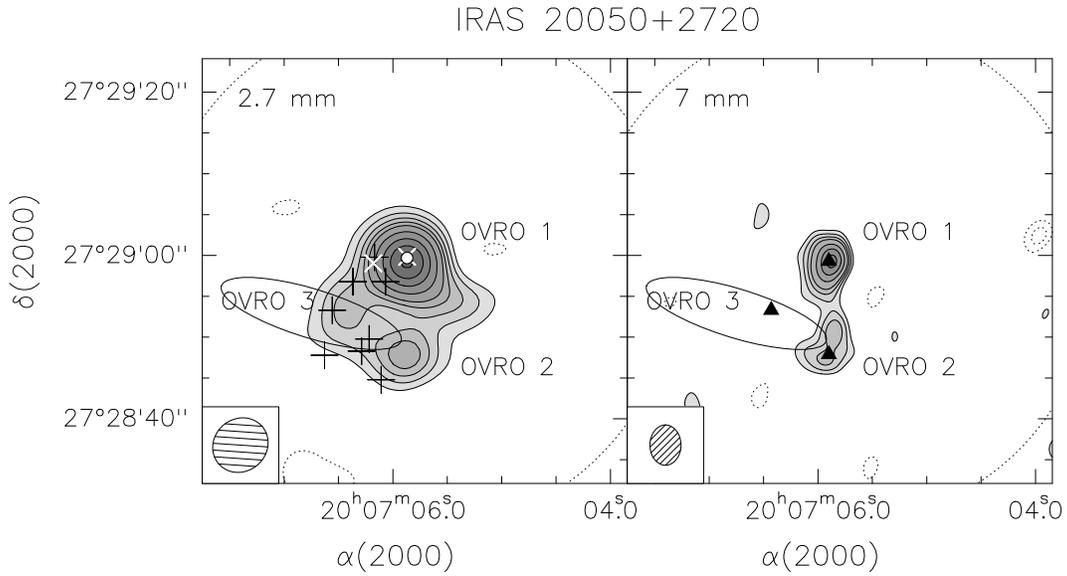}}
\caption{Map of the OVRO 2.7~mm ({\it left panel}) and the VLA 7~mm ({\it right panel}) continuum emission
towards  \iras. The rms noise of the map, $\sigma$, is 1.5~mJy\,beam$^{-1}$ ({\it left panel}) and 
0.16~mJy\,beam$^{-1}$ ({\it right panel}). The contour levels are $-5$, $-3$, 3, 5, 7, 9, 11, 15, 20, 25,
and 30 times $\sigma$ ({\it left panel}), and $-4$, $-3$, 3, 4, 5, 6, 8, 10, and 12 times $\sigma$ ({\it
right panel}). The synthesized beams are drawn in the bottom left corner. The error ellipse of \iras\ is
indicated in black. The dotted circles represent the OVRO ({\it left panel}) and VLA ({\it right panel})
primary beam (50\% attenuation level). Black crosses indicate the corrected position (see Sect. \ref{dust})
of the NIR sources detected by Chen et al.\ (\cite{chen97}), and white crosses the position of the 3~mm
sources detected by Furuya et al.\ (\cite{furuya05}). The white circle marks the position of the 3.6~cm VLA
source reported by Anglada et al.~(\cite{anglada98a}). Black triangles in  {\it right panel} mark the OVRO
2.7~mm continuum position of the sources.}
\label{cont}
\end{figure*}

\begin{table*}
\caption[] {Positions, millimeter flux densities, sizes, and masses of the cores detected towards \iras}
\label{table_clumps}
\begin{tabular}{l@{\hspace{-2pt}}c@{\hspace{-2pt}}c@{\hspace{-2pt}}cc@{\hspace{-2pt}}c@{\hspace{-2pt}}cc@{\hspace{-2pt}}c@{\hspace{-2pt}}c@{\hspace{-2pt}}c@{\hspace{-2pt}}c@{\hspace{-2pt}}c@{\hspace{-2pt}}c@{\hspace{-2pt}}c@{\hspace{-2pt}}c}
\hline
&\multicolumn{2}{c}{Position$^a$} 
&&\multicolumn{2}{c}{$\lambda=2.7$mm}
&&\multicolumn{2}{c}{$\lambda=7$mm} \\ 
 \cline{2-3} 
 \cline{5-6}  
 \cline{8-9}
&\multicolumn{1}{c}{$\alpha({\rm J2000})$} &
\multicolumn{1}{c}{$\delta({\rm J2000})$} &
&\multicolumn{1}{c}{$I^{\rm peak}_\nu$ $^{b}$} &
\multicolumn{1}{c}{$S_\nu$ $^{b}$} &
&\multicolumn{1}{c}{$I^{\rm peak}_\nu$ $^{~b}$} &
\multicolumn{1}{c}{$S_\nu$ $^b$} &
\multicolumn{1}{c}{$\theta^c$} &
\multicolumn{1}{c}{Mass$^d$} &
\multicolumn{1}{c}{$n({\rm H_2})^e$} &
\multicolumn{1}{c}{$N({\rm H_2})^e$} \\
\multicolumn{1}{c}{Core} &
\multicolumn{1}{c}{20$^{\rm h}$ 07$^{\rm m}$}&
\multicolumn{1}{c}{+27$^{\degr}$ 28$^{\arcmin}$} &
&\multicolumn{1}{c}{({mJy beam$^{-1}$})} & 
\multicolumn{1}{c}{(mJy)} &
&\multicolumn{1}{c}{({mJy beam$^{-1}$})} & 
\multicolumn{1}{c}{(mJy)} &
\multicolumn{1}{c}{(arcsec)} &
\multicolumn{1}{c}{($M_\odot$)} &
\multicolumn{1}{c}{($10^{6}$ cm$^{-3}$)} &
\multicolumn{1}{c}{(10$^{23}$ cm$^{-2}$)} \\
\hline
OVRO 1  &$05\fs90$ &$59\farcs3$ & &$\phantom{2} 46.4 \pm 9.3$ &$\phantom{11.} 65
\pm13\phantom{.1}$ & &$2.04\pm0.17$
&$2.88\pm0.17$ &3.3 &6.5 &1.3 &1.4\\
OVRO 2  &$05\fs90$ &$47\farcs9$ & &$\phantom{1} 17.1\pm3.4$ &$\phantom{1}
18.0\pm3.6\phantom{0}$ &&$0.97\pm0.17$ &$1.84\pm0.17$ 
&2.3 &$1.8$
&1.7 &1.1  \\ 
OVRO 3  &$06\fs43$ &$53\farcs3$ &&$\phantom{1} 12.9\pm2.6$ &$12.9\pm2.6$  &&$<0.48^{f}$ &$\ldots$ 
&$<6.7^{g}\phantom{1}$ &$\phantom{.}1.3^{h}$
&$\phantom{.}3.0^{h}$ &$\phantom{.}1.4^{h}$  
\\ 
\hline

\end{tabular}
  (a) Position of the 2.7~mm emission peak. \\
  (b) Corrected for primary beam response.  \\
  (c) Deconvolved geometrical mean of the major and minor axes of the half-peak intensity contour. \\  
  (d) Circumstellar mass estimated taking into account all the emission inside the 3~$\sigma$
  level, and assuming a dust temperature of 34~K (see Sect. \ref{mass}). \\
  (e) Average H$_2$ volume and column density estimated assuming spherical symmetry 
  and a $10\%$ Helium abundance. The densities have been computed taking into account the total
  circumstellar mass and the size of the source at a 3$~\sigma$ emission level. \\
  (f) 3~$\sigma$ upper limit. \\
  (g) Source not resolved. \\
  (h) Estimated from the peak intensity. \\
  \end{table*}

\section{Results}

\subsection{Dust emission}
\label{dust}

The  maps of the millimeter continuum emission around the embedded source \iras\  at 2.7 and
7~mm  are shown in Fig.~\ref{cont}. The continuum emission at 2.7~mm has been resolved into
three different sources, labeled OVRO~1, OVRO~2, and OVRO~3. OVRO~1 and OVRO~2 have also been
detected at 7~mm, but not OVRO~3, for which an upper limit (3~$\sigma$) of 0.48~mJy\,beam$^{-1}$
has been estimated. OVRO~3 is not so well defined at 2.7~mm as the other two sources. However,
the source is located within $\sim$$1\farcs6$ of the near-infrared (NIR) source 5 of Chen et
al.~(\cite{chen97}; see discussion below about the correction to the positions given by these
authors). The source is probably associated with \iras\ (Fig.~\ref{cont}), at least at 12 and
25~$\mu$m wavelengths. However, one cannot discard the possibility that part of the emission
detected at longer wavelengths (60 and 100~$\mu$m) is associated with OVRO~1 and/or OVRO~2 as
suggested by the IRAS high-resolution processed (HIRES\footnote{HIRES images are available as
single fields ($1\degr$--$2\degr$) through the Infrared Processing and Analysis Center (IPAC).})
images. OVRO~3 is also associated with a C$^{18}$O peak (see Sect.~\ref{c18o}). All this lead us
to consider the bump visible at 2.7~mm emission as an independent source, that we called OVRO~3.
OVRO~1 and OVRO~2 are separated  $\sim$$11\arcsec$ ($\sim$8000~AU at the distance of the
region), while OVRO~3 is separated $\sim$$9\arcsec$ ($\sim$6000~AU) from the other two sources.
The position, the flux density at 2.7~mm, the deconvolved size of the sources (measured as the
geometrical mean of the major and minor axis of the half-peak intensity contour around each
source), the mass, and the average volume and column density  are given in
Table~\ref{table_clumps}. 

\begin{figure*}
\centerline{\includegraphics[angle=-90,width=17cm]{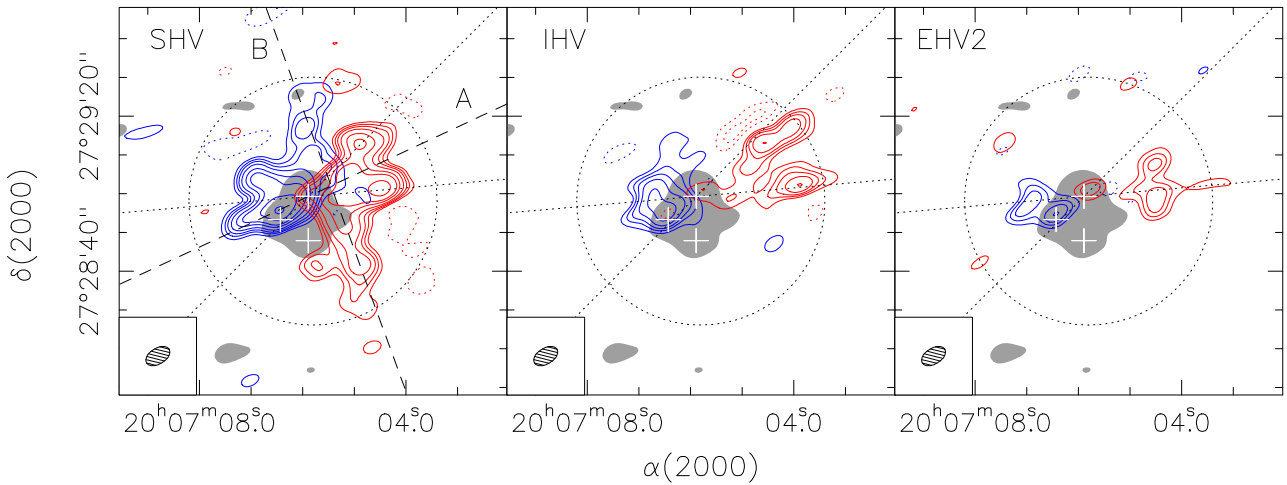}}
\caption{CO (\juz) emission integrated in different velocity intervals, ($-7$, +$1$) \kms\  for the
SHV blueshifted emission ({\it left panel; blue contours}), ($11$, $19$) \kms\ for the
redshifted one ({\it left panel; red contours}), ($-14$, $-7$) \kms\ for the IHV
blueshifted emission ({\it middle panel; blue contours}), and (+$19$, +$26$) \kms\, for the redshifted
one ({\it middle panel; red contours}), ($-19$, $-14$) \kms\ for the EHV2 blueshifted emission
({\it right panel; blue contours}), and (+$26$, +$31$) \kms\, for the redshifted one ({\it right panel;
red contours}), overlaid on the 2.7~mm continuum emission at a 3~$\sigma$ level {\it(greyscale)}.
Contour levels are $-5$, $-3$, 3, 5, 7, 10, 15, 20, 30 and 40 times $\sigma$, where 1~$\sigma$ is
0.6\jykms\ for the SHV interval, 0.5\jykms\ for IHV, and 0.25\jykms\ for EHV2. The
white crosses mark the millimeter continuum positions of OVRO~1,  OVRO~2 and OVRO~3. The black dashed
lines in {\it left panel} outline the direction of outflows A, and B (see Sect.~\ref{co}). The black dotted lines outline
the edges of outflow A powered by OVRO~1.  The synthesized beam is drawn in the bottom left 
corner. The dotted circle represents the OVRO primary beam (50\% attenuation level).}
\label{co_outflow} 
\end{figure*}

The strongest millimeter source is OVRO~1. This source is located at the center of one of the two CO
molecular outflows detected in the region (see the next section and Fig.~\ref{co_outflow}), which
suggests that OVRO~1 is its driving source. OVRO~1 has a core-halo structure at millimeter wavelengths,
with a core that has a deconvolved diameter of $\sim$$3\farcs3$ ($\sim$2300 AU) at the half-peak
intensity contour of the dust emission, and an extended and quite spherical halo or envelope that has a
size of $\sim$$11\farcs3$ ($\sim$8000 AU) at the 3~$\sigma$ contour level. The source OVRO~2 has a
deconvolved diameter of $\sim$$2\farcs3$ ($\sim$1600 AU) at the half-peak intensity contour of the dust
emission. On the other hand, OVRO~3 is not resolved with the angular resolution of the 2.7~mm
observations. These sizes are consistent with the values found for envelopes around low- and
intermediate-mass protostars (e.g., Hogerheijde et al~\cite{hogerheijde97}, \cite{hogerheijde99};
Looney et al.~\cite{looney00}; Fuente et al.~\cite{fuente01}; Paper I, Paper II). As can be seen in
Fig.~\ref{cont}, there is extended emission surrounding the three sources, which is clearly visible at
2.7~mm but not at 7~mm. The estimated upper limit (3~$\sigma$) for the extended emission at 7~mm
is 0.48~mJy\,beam$^{-1}$.  This extended component has a size of $\sim$$19''$ ($\sim$13000 AU). 

 The position of the eight NIR sources detected by Chen et al.~(\cite{chen97}) are indicated as black crosses in
Fig.~\ref{cont}. We compared these positions with those of the 2MASS catalog and found that the positions reported
by Chen et al.~(\cite{chen97}) were systematically offset on average by $\sim$$ - 0\fs29$ in right ascension, and
$\sim$$1\farcs4$ in declination. Therefore, before plotting them, we corrected the positions. As already
mentioned, the only NIR source possibly associated  with the millimeter sources detected with OVRO  at 2.7~mm is
source~5, whose position nearly coincides with OVRO~3. Chen et al.~(\cite{chen97}) also detect a compact source at
2.7~mm with the Berkeley-Illinois-Maryland Association (BIMA) array coincident with  OVRO~1. These authors estimate
a flux density of $30\pm3.5$~mJy, which is half of that derived with OVRO (see Table~\ref{table_clumps}). Although
Chen et al.~(\cite{chen97}) locate the millimeter continuum source close to the NIR sources 6, 7, and 8, this is
not the case after correcting the infrared positions. Therefore, this suggests that OVRO~1 and OVRO~2 could be
deeply embedded sources still undetected at NIR wavelengths. The situation could be different for OVRO~3 because of
its possible association with \iras\ and NIR source~5. 

\begin{figure*}
\centerline{\includegraphics[angle=-90,width=14cm]{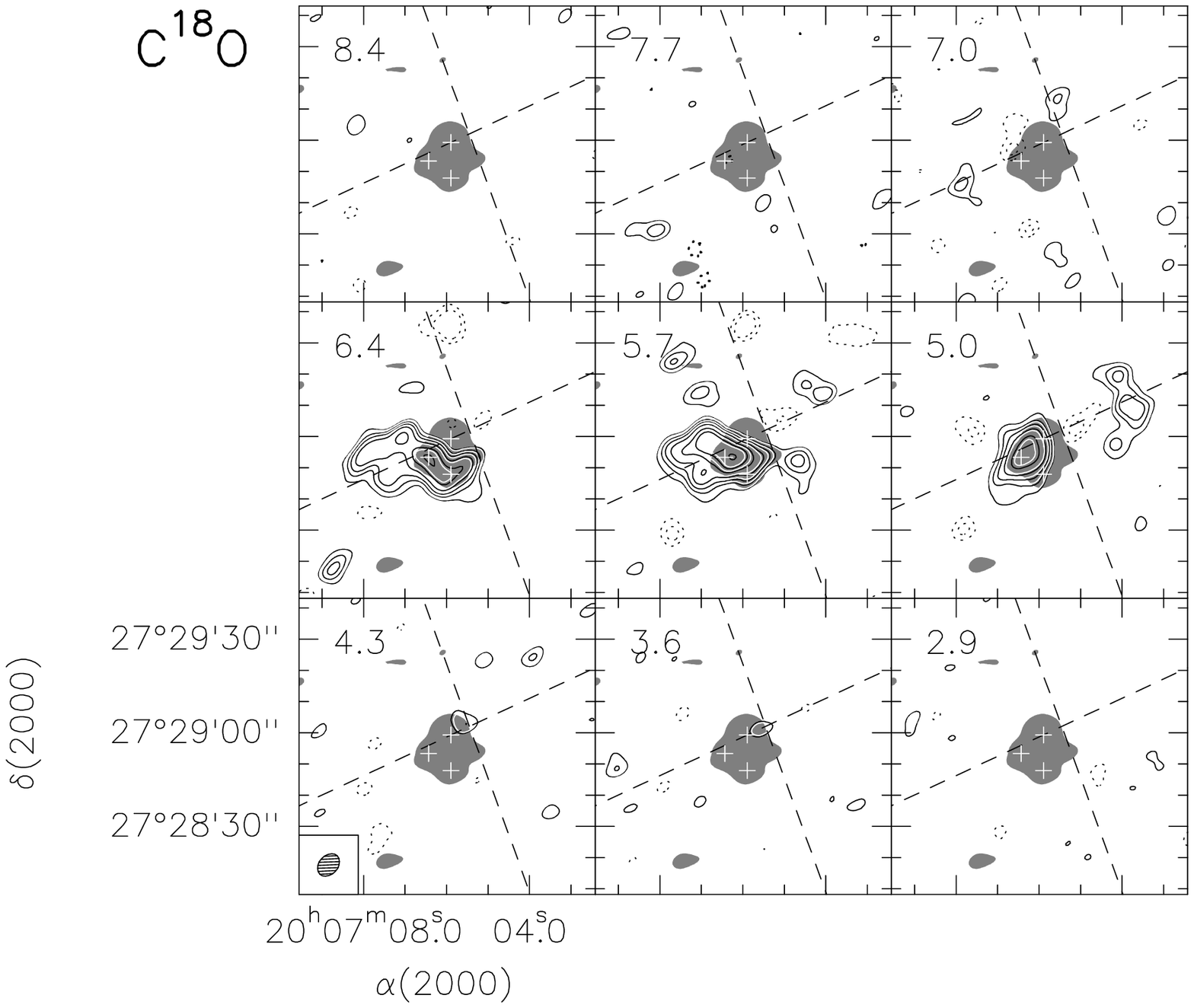}}
\caption{Velocity channel maps of the C$^{18}$O~(\juz) emission {\it (contours)} overlaid
on the 2.7~mm continuum emission  at a 3~$\sigma$ level {\it (greyscale)}. The systemic velocity of \iras\ is
$6$~\kms.  The central velocity of each channel is indicated at the upper left corner of
the panels.  The 1~$\sigma$ noise in one channel is 40~mJy\,beam$^{-1}$.  Contour levels are
$-5$, $-3$, 3, 5, 7, 10, 15, 20, and 30 times $\sigma$. The conversion factor from
Jy\,beam$^{-1}$ to K is 2.10. The synthesized beam is drawn in the bottom left  corner of
the bottom left panel. The white crosses mark the millimeter continuum positions of OVRO~1, OVRO~2, and
OVRO~3. The dashed lines outline the direction of the CO outflows A and B detected in the region (see
Fig.~\ref{co_outflow}).} 
\label{c18o_channel}
\end{figure*}

The region has also been observed at centimeter wavelengths with the VLA by Anglada et
al.~(\cite{anglada98a}). As can be seen in Fig.~\ref{cont}, the source OVRO~1 coincides within $0\farcs5$
with a 3.6~cm source, with two components separated by $\sim0\farcs3$ (Anglada et al.~\cite{anglada98a}).

The position of the millimeter source detected by Furuya et al.~(\cite{furuya05}) with the OVRO
interferometer at 3~mm is indicated as a white cross in Fig.~\ref{cont}. The source is clearly
coincident with OVRO~1. These authors also find millimeter emission eastward of OVRO~1, which
could be associated with one of the NIR sources. This suggests that another YSO could be
embedded in the core (Furuya et al.~\cite{furuya05}). The position of this possible embedded
YSO detected by Furuya et al.~(\cite{furuya05}) is also marked in Fig.~\ref{cont} with a white
cross. Our observations do not have enough resolution to separate this eastern emission from
that of OVRO~1 and its extended envelope. The total integrated flux density measured by Furuya
et al.~(\cite{furuya05}) at 3~mm is 25.6~mJy. If we assume a dust absorption coefficient
proportional to $\nu^2$, the expected flux at 2.7~mm would be $\sim 39$~mJy, which is about
half of the integrated flux density measured for OVRO~1 at 2.7~mm. This is probably due to the
fact that the Furuya et al.\ observations were carried out with more extended OVRO
configurations (E and H), and
thus, being somewhat less sensitive to extended emission. As a consequence of this, it is possible that
part of the extended emission surrounding OVRO~1 was filtered out by the interferometer,
consistent with the negative contours visible in Fig.~1 of Furuya et al.~(\cite{furuya05}).

Choi et al.~(\cite{choi99}) observed the region with BIMA at 3.4~mm, with a synthesized beam slightly
larger than that of our 2.7~mm OVRO observations, and resolve the continuum emission into four
sources, labeled A, B, C, and D. The peak position of source A is in agreement with that of OVRO~1.
This source is extended and elongated towards the south and the southeast, and its structure resembles
that of the extended emission surrounding OVRO~1, 2, and 3 as mapped by us at 2.7~mm. 

Chini et al.~(\cite{chini01}) observed the region with the single dish James Clerk Maxwell Telescope (JCMT) at
0.45, 0.85, and 1.3~mm with angular resolutions ranging from $\sim$8 to 15$''$ and detected three sources in the region, \iras\ MMS~1, MMS~2, and MMS~3, together
with an extension to the south of MMS~1. The source \iras\ MMS~1 is associated with OVRO~1, 2 and 3.
Beltr\'an et al.~(\cite{beltran06a}) detect a similar emission with the Swedish-ESO Submillimetre
Telescope (SEST) at 1.3~mm. The clump number 1 is the one associated with OVRO~1, 2, and 3. However, this
clump, which has a size of $58\farcs2$, is much larger than the extended 2.7~mm emission or MMS~1, as
it includes the extended southern emission as well.

\begin{table*}
\caption[] {C$^{18}$O (\juz), and HC$_3$N (\jdo) line and physical parameters towards 
 OVRO~1, OVRO~2 and OVRO~3 in \iras}
\label{table_lines}
\begin{tabular}{llcccccc}
\hline
 &
 &
\multicolumn{1}{c}{\Vlsr} &
\multicolumn{1}{c}{$\Delta V$}  &
\multicolumn{1}{c}{$T_{\rm B}$} &
\multicolumn{1}{c}{$\int{T_{\rm B}\,{\rm d}V}$} &
\multicolumn{1}{c}{$\theta^a$} &
\multicolumn{1}{c}{$M_{\rm vir}^b$}
\\
\multicolumn{1}{c}{Position} &
\multicolumn{1}{c}{Molecule} &
\multicolumn{1}{c}{(km s$^{-1}$)} &
\multicolumn{1}{c}{(km s$^{-1}$)} &
\multicolumn{1}{c}{(K)} &
\multicolumn{1}{c}{(K km s$^{-1}$)} &
\multicolumn{1}{c}{(arcsec)} &
\multicolumn{1}{c}{($M_\odot$)} 
\\
\hline
OVRO 1 &C$^{18}$O  &$4.68\pm0.04$ &$1.50\pm0.21$ &$0.94\pm0.11$ &$1.50\pm0.07$ & &\\
       &HC$_3$N  &$4.82\pm0.02$ &$1.44\pm0.05$ &$3.80\pm0.12$ &$5.84\pm0.18$ &$9\farcs6$
       &4.3--5.7\\
\hline
OVRO 2 &C$^{18}$O  &$6.33\pm0.04$ &$1.70\pm0.10$ &$1.56\pm0.19$ &$2.83\pm0.12$ &$7\farcs0^{c}$
&$<$4.3--5.8\phantom{2}\\
       &HC$_3$N  &$5.86\pm0.08$ &$1.60\pm0.20$ &$0.85\pm0.13$ &$1.44\pm0.17$ &$$ &\\
\hline
OVRO 3 &C$^{18}$O  &$5.61\pm0.03$ &$1.92\pm0.06$ &$2.51\pm0.17$ &$5.14\pm0.15$ &$14$
&11--15 \\
               &HC$_3$N  &$5.41\pm0.03$ &$1.30\pm0.10$ &$2.06\pm0.14$ &$2.85\pm0.14$ &$$ &\\
\hline 
\end{tabular}

(a) Deconvolved geometrical mean of the major and minor axes of the half-peak intensity contour. \\
(b) Estimated from Eq.~5 of Paper~II for typical density distributions with $p=2.0$--1.5. \\
(c) Emission not resolved. Beam size. 
\end{table*}

\subsection{The molecular outflow: CO emission}
\label{co}

The CO emission towards \iras\ has been previously studied through lower
angular resolution observations by Bachiller et al.\ (\cite{bachiller95}). These
authors discover three bipolar molecular outflows in the region:  an
extremely high velocity highly collimated outflow oriented nearly east-west
(labeled A in Fig.~3 of Bachiller et al.); a bipolar outflow of intermediate
high velocity oriented in the northeast-southwest direction (labeled B); and a
standard high velocity bipolar outflow detected along the northwest-southeast
direction (labeled C). The observations reported in the present study significantly 
improve the angular resolution of previous outflow maps, revealing the
structure of the molecular outflows in detail, as the interferometer filters out
most of the extended CO emission seen in the Bachiller et
al.~(\cite{bachiller95}) low and intermediate velocity maps.

\begin{figure}
\centerline{\includegraphics[angle=0,width=7.5cm]{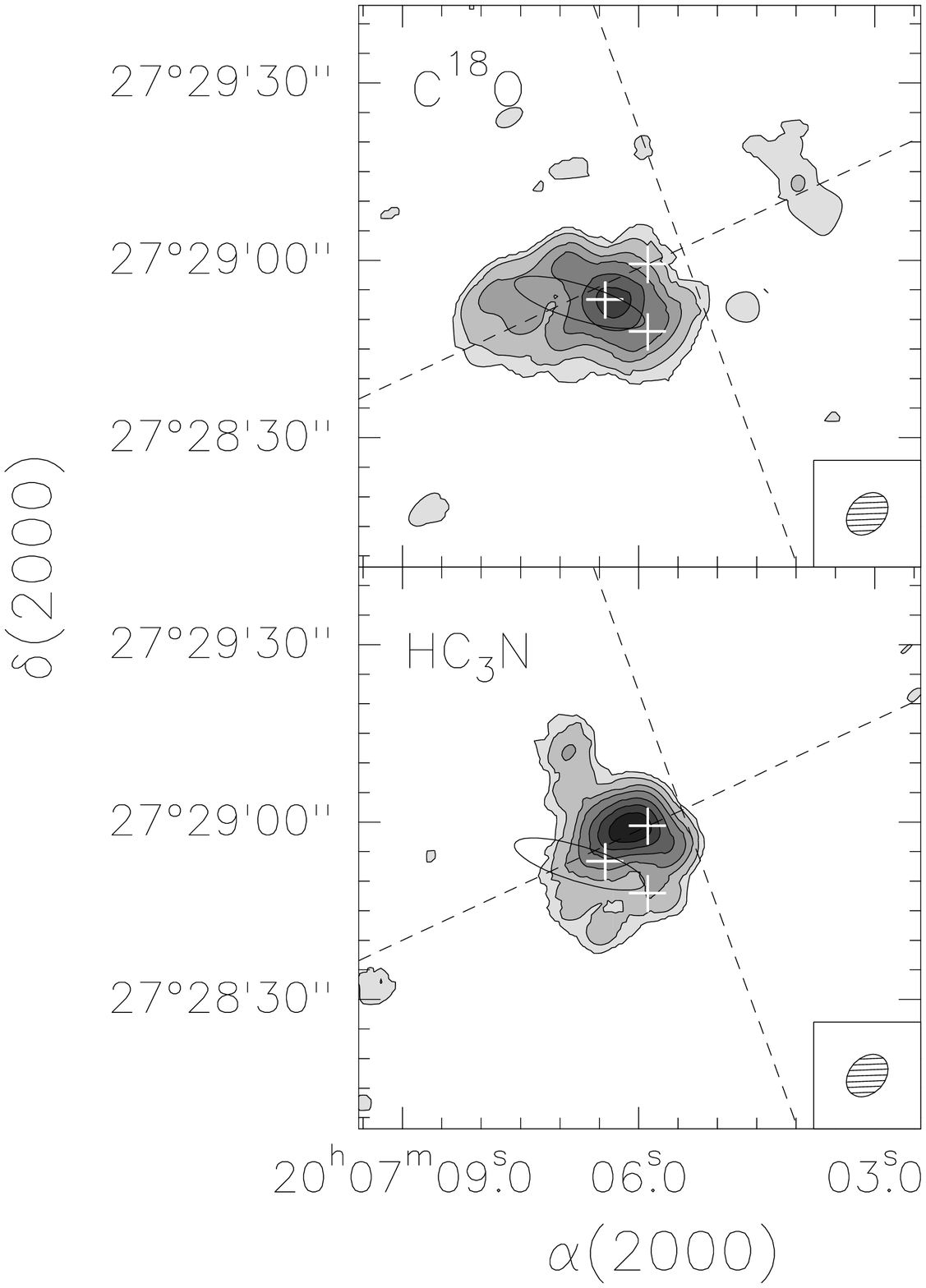}}
\caption{Integrated intensity map of the C$^{18}$O (\juz) ({\it top}) and the
HC$_3$N~(\jdo)  ({\it bottom}) emission over the velocity interval (3,8)~\kms\ towards
\iras.  Contour levels are 0.17, 0.34, 0.68, 1.0, 1.5, 2.0, and 2.5\jykms\ for both panels. 
The white crosses show the millimeter continuum positions of OVRO~1, OVRO~2, and OVRO~3. The dashed lines indicate the
direction of the CO outflows A and B. The error ellipse of \iras\ is indicated in black. The
synthesized beam is drawn in the bottom right corner of each panel.}
\label{average}
\end{figure}

Figure~\ref{co_outflow} shows the maps of the integrated CO (\juz) emission in three different blueshifted
and redshifted velocity intervals. For the sake of comparison, we chose the same velocity intervals as
Bachiller et al.~(\cite{bachiller95}): a standard-high velocities (SHV) interval, with ($-7$, +$1$)~\kms\
for blueshifted emission and ($11$, $19$)~\kms\ for the redshifted one;  an intermediate-high velocities
(IHV) interval, with ($-14$, $-7$)~\kms\ for the blueshifted emission and  (+$19$, +$26$)~\kms\, for the
redshifted one; and an extremely-high velocities (EHV2) interval, with ($-19$, $-14$)~\kms\ for the
blueshifted emission and (+$26$, +$31$)~\kms\, for the redshifted one. The other extremely high-velocities
interval, EHV1, observed by Bachiller et al.~(\cite{bachiller95}) has not been mapped because our OVRO
observations did not cover that range of velocities. The systemic velocity, $V_{\rm LSR}$, is roughly
6~\kms. As can be seen in this figure, the CO emission traces two bipolar outflows in the region, a roughly
east-west outflow, coincident with outflow A of Bachiller et al.~(\cite{bachiller95}), and a
northeast-southwest outflow, coincident with outflow B. The outflow labeled C by Bachiller et
al.~(\cite{bachiller95}) and mapped at the lowest velocity interval (SHV) has not been detected. As can be
seen in Fig.~3 of Bachiller et al.~(\cite{bachiller95}), the outflow C, which has a PA of about
$-35\degr$,   has very extended lobes, especially for the blueshifted emission that reaches distances of
$>$$100''$ from the IRAS source. Most of the emission of this outflow lies outside of the OVRO primary beam
and, therefore, these interferometric observations are not sensitive to it. Note that one cannot rule out
the possibility that part of the central emission of the outflow C overlaps that of the outflows A and B,
and as it is not possible to distinguish any possible emission of the outflow C from that of the outflows A
and B, we considered the outflow C as not detected in this study.

\subsection{C$^{18}$O emission}
\label{c18o}

\begin{figure}
\centerline{\includegraphics[angle=-90,width=8.8cm]{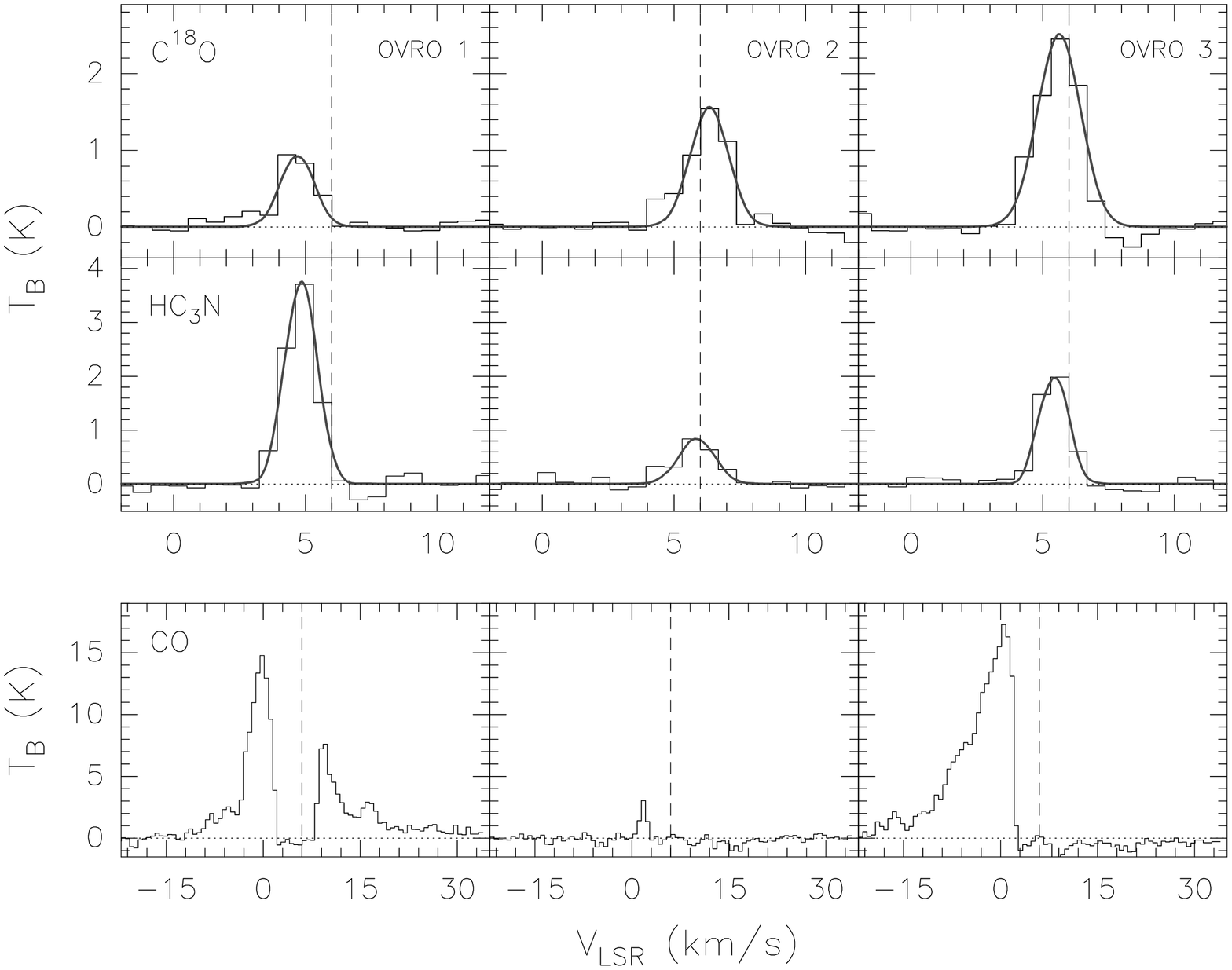}}
\caption{C$^{18}$O (\juz) ({\it top}), HC$_3$N (\jdo) ({\it middle}), and CO (\juz) ({\it
bottom}) spectra obtained at the position of OVRO~1, OVRO~2, and OVRO~3 in
\iras.  The continuum has been subtracted.
The conversion factor is 2.10~K/Jy beam$^{-1}$ for C$^{18}$O, 2.14~K/Jy beam$^{-1}$ for
HC$_3$N, and 3.30~K/Jy beam$^{-1}$ for CO. The thick grey profiles in {\it top} and {\it middle} panels are the Gaussian fits to the spectra.  The dashed vertical
line indicates the systemic velocity of 6~\kms.}
\label{lines}
\end{figure}

Figure~\ref{c18o_channel} shows the velocity channel maps for the  C$^{18}$O~(\juz) emission around the
systemic velocity, $V_{\rm LSR}\simeq 6$~\kms, overlaid on the continuum emission. The emission integrated
over the central channels, velocity interval (3, 8)~\kms, is shown in the top panel of Fig.~\ref{average}. 
The emission is elongated in the east-west direction and peaks towards OVRO~3. C$^{18}$O shows a more
extended emission than the dust. In particular, it shows an elongation in the direction of the molecular
outflow A, which suggests some interaction between the outflow and the envelope. The spectra taken at the
position of the continuum peak of OVRO~1, OVRO~2, and OVRO~3 are shown in Figure~\ref{lines}.
Table~\ref{table_lines} lists the fitted parameters for C$^{18}$O (\juz) and HC$_3$N (\jdo). The
corresponding Gaussian fits are shown in Fig.~\ref{lines}. As can be seen in the channel maps and the
spectra, the C$^{18}$O emission is associated with the three millimeter sources, with the emission towards
OVRO~1 weaker than towards OVRO~2 and 3. The velocity integrated flux density of the central emission is
$\sim$16~Jy\,\kms\ and  has a deconvolved size, measured as the geometrical mean of the major and minor
axes of the half-peak intensity contour of the gas, of $\sim$$14''$ or $\sim$9800~AU.

\begin{figure*}
\centerline{\includegraphics[angle=-90,width=14cm]{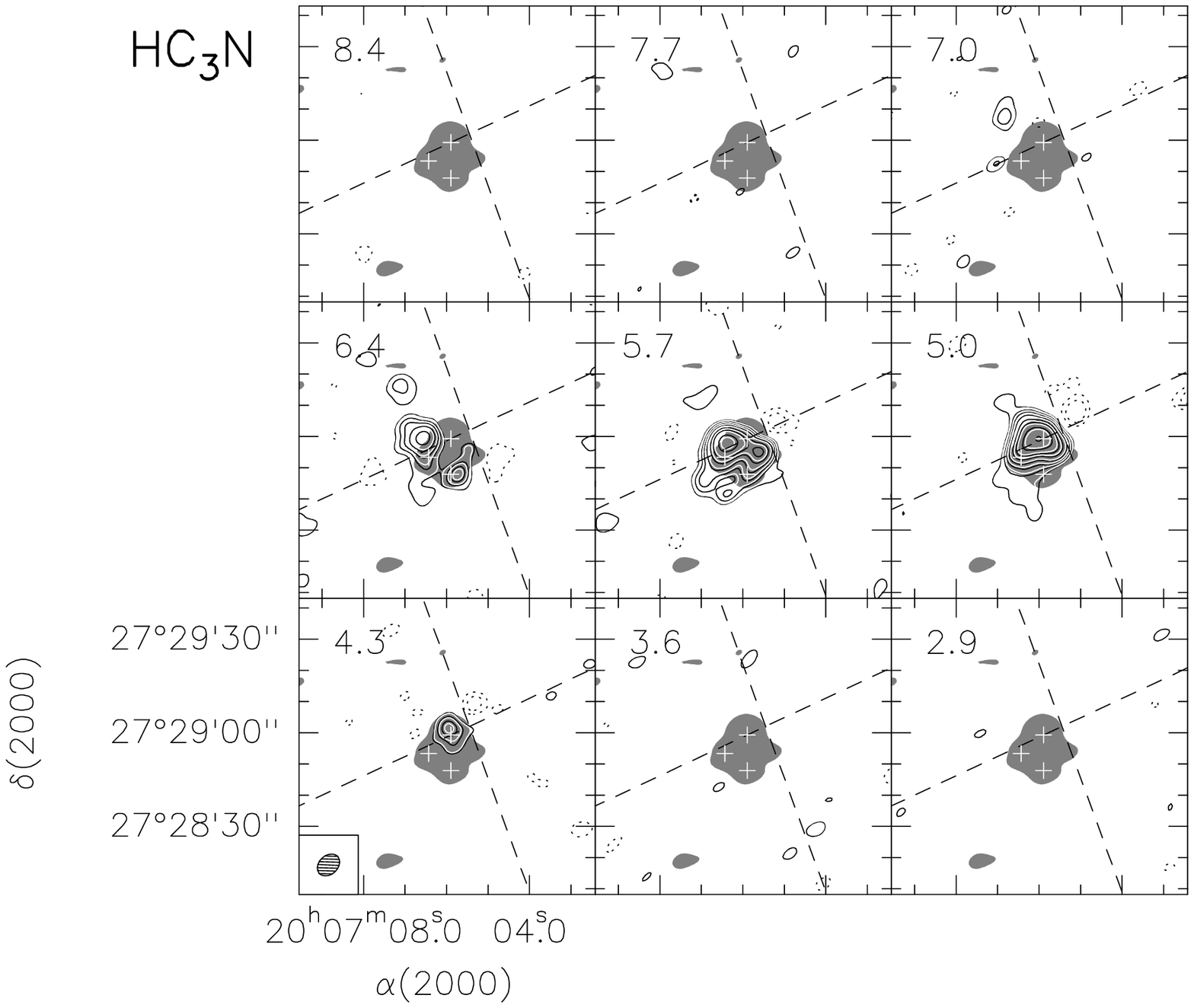}}
\caption{Velocity channel maps of the HC$_3$N~(\jdo) emission {\it (contours)} overlaid on
the 2.7~mm continuum emission  at a 3~$\sigma$ level {\it (greyscale)}. The systemic velocity of \iras\ is
$6$~\kms.  The central velocity of each channel is indicated at the upper left corner of
the panels. The 1~$\sigma$ noise in one channel is 40~mJy\,beam$^{-1}$.  Contour levels are
$-5$, $-3$, 3, 5, 7, 10, 15, 20, 30 and 45 times $\sigma$. The conversion factor from Jy\,beam$^{-1}$ to K is
2.14. The synthesized beam is drawn in the bottom left  corner of the bottom left panel.
The white crosses mark the millimeter continuum positions of OVRO~1, OVRO~2, and OVRO~3. The dashed lines
outline the direction of the CO outflows A nd B (see Fig.~\ref{co_outflow}).}
\label{hc3n_channel}
\end{figure*}

\subsection{HC$_3$N emission}
\label{hc3n}

Figure~\ref{hc3n_channel} shows the velocity channel maps for the HC$_3$N~(\jdo) emission towards \iras,
around the systemic velocity, $V_{\rm LSR}\simeq 6$~\kms, overlaid on the continuum emission. The emission
integrated over the velocity interval (3, 8)~\kms\ is shown in the bottom panel of Fig.~\ref{average}. 
HC$_3$N shows a spatial distribution different from that of C$^{18}$O and the dust. 
This is probably because the different molecular tracers and the dust are tracing material at slightly
different physical conditions. The C$^{18}$O elongation along outflow A is not traced by HC$_3$N, which could
indicate that the gas is not dense enough to excite the HC$_3$N~(\jdo) transition. The spectra taken at the position of the
continuum peak of OVRO~1, OVRO~2, and OVRO~3 are shown in Figure~\ref{lines}. The emission
peaks about $4\farcs2$ east of the continuum peak position of OVRO~1, and it is clearly
associated with this source.  HC$_3$N  emission is also associated with the other two
sources in the region, with the emission towards OVRO~2 weaker than towards OVRO~1 and 3.
Although the emission is quite compact, it shows a tail towards the northeast, and it is
slightly elongated towards the southeast, along the direction of the molecular outflow A.
The HC$_3$N emission has a velocity integrated flux density of $\sim$12.7~Jy\,\kms\ and a deconvolved size
of $9\farcs6$ or $\sim 6700$~AU.

\section{Analysis and discussion of IRAS~20050+2720}

\subsection{Mass and density estimates from dust emission}
\label{mass}

Assuming that the dust emission is optically thin, we estimated the masses of the sources
(Table~\ref{table_clumps}) following Eq.~1 of Paper~II and adopting a dust mass opacity
coefficient at 111~GHz, $\kappa_{111}$=0.2~cm$^2$\,g$^{-1}$ ($\kappa_{0}$ = 1~cm$^2$\,g$^{-1}$
at 250~GHz: Ossenkopf \& Henning~\cite{ossenkopf94}) and a gas-to-dust ratio of 100.  The dust
temperature assumed to derive the masses is 34~K, which is the dust temperature obtained by
Chini et al.~(\cite{chini01}) by fitting a grey-body to the IRAS fluxes at 60 and 100~$\mu$m and
the SCUBA fluxes at 450, 850, and 1300~$\mu$m.  The mass estimated is 6.5~$M_{\odot}$ for
OVRO~1, 1.8~$M_{\odot}$ for OVRO~2, and  1.3~$M_{\odot}$ for OVRO~3. Note that using a
lower dust mass opacity coefficient as proposed by Hildebrand~(\cite{hildebrand83}) would result
in masses about a factor 4 higher, whereas temperatures 10~K lower would increase the masses by
about a factor 1.5. The circumstellar masses of the three deeply embedded sources are similar
to those found for intermediate-mass YSOs for OVRO~1, and for low-mass protostars for OVRO~2 and
3. The average H$_2$ volume density, $n({\rm H_2})$, and column density, $N({\rm H_2})$, of the
sources estimated by assuming spherical symmetry are given in Table~\ref{table_clumps}.

\subsection{The molecular outflows}

\subsubsection{Morphology}
\label{co_morph}

\begin{figure}
\centerline{\includegraphics[angle=-90,width=8.8cm]{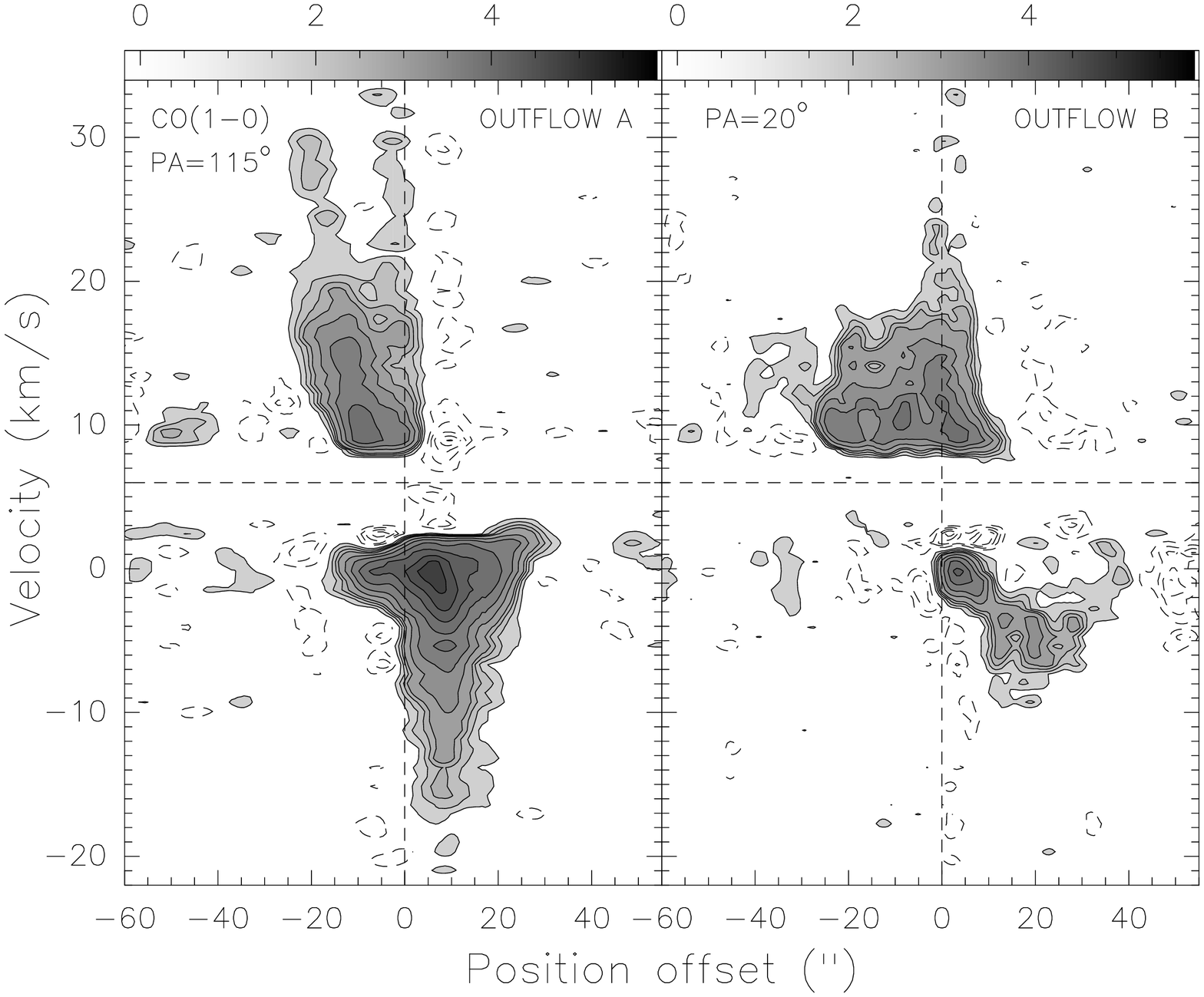}}
\caption{PV plot of the CO (\juz) emission along the major axis of the molecular
outflow A with PA$=115\degr$ ({\it left}), and of the outflow B with
PA$=20\degr$ ({\it right}). The position offset is relative to the position of
OVRO 1 ({\it left}) and relative to the center of symmetry of the outflow B, $\alpha$(J2000) = $20^{\rm h}
07^{\rm h} 05\fs38$,  $\delta$(J2000) = $+27\degr 28'  59\farcs3$. Contours are $-1.2$, $-0.9$, $-0.72$, $-0.54$, $-0.36$,
$-0.18$, 0.18, 0.36, 0.54, 0.72, 0.90, 1.2, 1.8, 2.7, 3.6, 4.5, and 5.4~Jy\,beam$^{-1}$. The horizontal line marks the systemic velocity, $V_{\rm
LSR}=6$~\kms.}
\label{co_cut}
\end{figure}

The outflow A, which has been detected at the three velocity intervals, is centered at the position of
OVRO~1. It is elongated in the direction of PA $\simeq$$115\degr$, and presents a different morphology for
the blueshifted and the redshifted emission. The blueshifted emission is more compact than the redshifted
one. This lack of symmetry between the lobes could be due to the presence of OVRO~3. In such a scenario,
the redshifted gas would flow freely westwards of the powering source OVRO~1, while the blueshifted gas,
moving eastwards, would interact with the OVRO~3 core that in projection seems to be located in the
direction of the blue lobe. The CO spectra towards OVRO~3 in Fig.~\ref{lines} exhibits a high-velocity
blueshifted wing which supports the ''impact'' hypothesis. Such a situation resembles that in
IRAS~21391+5802 (Paper~I), where the molecular outflow interacts with the dense material in the
surroundings and gets deflected. The blueshifted emission shows a lobe that shrinks as the velocity of the
outflow increases, and the direction of the emission is roughly the same for all the velocity intervals. On
the other hand, the redshifted emission is more extended and shows a more complex morphology. At standard
(SHV) and intermediate (IHV) outflow velocities, the emission seems to trace the walls of the cavity
excavated by the outflow. As the velocity of the outflow increases, the angle of aperture of the conical
cavity decreases, and the redshifted emission is detected at larger distances from OVRO~1. The farthest
redshifted emission is located at $\sim$$35''$ from OVRO~1 at the IHV interval. CO is detected at the
center of the lobe only at EHV2. In fact, at EHV2, the redshifted emission is not tracing the walls of the
cavity. The fact that no CO emission is detected inside the cavity for SHV and IHV indicates that the gas
has already been evacuated. This suggests that the outflow and therefore the powering source OVRO~1 could
be in a later Class~0 evolutionary stage. Figure~\ref{co_cut} shows the position-velocity (PV) plot along
the major axis of the outflow A and centered at the position of OVRO~1,  and the PV plot along the major
axis of the outflow B and centered at the center of symmetry of the outflow (see below). As can be seen in
Fig.~\ref{co_cut}, for the outflow A there is blueshifted emission at negative offsets, and for the outflow
B redshifted emission at positive offsets. This probably indicates that the two outflows are overlapped along
the line-of-sight near their centers of symmetry where the powering sources are embedded or that the
outflows are not resolved at the base by the current resolution.

The bipolar outflow B has a PA of $\sim$$20\degr$ and it has been clearly detected at SHV. At IHV,
only some weak blueshifted emission towards the NE is visible. The powering source is
probably located close to the center of symmetry of the lobes. However, although there is some extended
2.7~mm continuum emission towards that position, the angular resolution of our observations is not high enough to
separate the emission of the source driving outflow B from that of the circumstellar envelope
surrounding OVRO~1. There are several 2MASS
infrared sources close to the center of symmetry of the outflow, although it is difficult to determine
whether one of these could be the driving source. The emission of this outflow  presents two
well-defined and elongated lobes, which have an extension of $\sim$$30''$.   

\subsubsection{Physical parameters of the CO outflows}
\label{co_prop}

\begin{table*}
\caption[] {Properties of the CO outflow}
\label{tco}
\begin{tabular}{lccccc}
\hline
&\multicolumn{1}{c}{$V^{a}$}&
\multicolumn{1}{c}{$M^{b}$}&
\multicolumn{1}{c}{$P^{c}$}&
\multicolumn{1}{c}{$E^{c}$}&
\multicolumn{1}{c}{$F_{\rm out} ^{c}$}
\\
\multicolumn{1}{c}{Lobe}&
\multicolumn{1}{c}{(\kms)}&
\multicolumn{1}{c}{($M_\odot$)}&
\multicolumn{1}{c}{($M_\odot$ \kms)}&
\multicolumn{1}{c}{($10^{44}$ erg)}&
\multicolumn{1}{c}{($M_\odot \, \mbox{km s}^{-1}$\,yr$^{-1}$)}
\\
\hline
&&&  OUTFLOW A \\
\hline
Blue &[$-19, +1$]          &0.10  &0.99  &1.05  &$2.8\times10^{-4}$\\
Red &[$+11, +31$]          &0.10  &1.05  &1.32  &$2.5\times10^{-4}$ \\
Total &                    &0.20  &2.04  &2.37  &$5.3\times10^{-4}$ \\
\hline
&&&  OUTFLOW B \\
\hline
Blue &[$-14, +1$]          &0.02  &0.23  &0.24  &$6.4\times10^{-5}$\\
Red &[$+11, +19$]          &0.03  &0.19  &0.14  &$4.5\times10^{-5}$ \\
Total &                    &0.05  &0.42  &0.38  &$1.1\times10^{-4}$ \\
\hline

\end{tabular}

(a)  Range of outflow velocities.  \\
(b) Assuming an excitation temperature of 29~K. \\
(c) Momenta and kinetic energies are calculated relative to the cloud velocity, which is taken to be $V_{\rm LSR}=6$~\kms. Velocity not corrected for inclination. \\

\end{table*}

Assuming that the CO emission is in LTE and is optically thin, the mass of the gas associated with
the outflows detected in the region has been calculated by using Eq.~2 of Paper~II. The high
outflow velocities CO is expected to be optically thin (e.g.\ Arce \& Goodman \cite{arce01}). If CO
is optically thick in portions of the flow, the correction for opacity should be in any case small .
As the CO spectra are complex and exhibit prominent self-absorption in both interferometric (see
Fig.~\ref{lines}) and single-dish observations (Bachiller et al.~\cite{bachiller95}; Zhang et
al.~\cite{zhang05}), it is difficult to derive the excitation temperature from the CO brightness
temperature. Therefore, taking into account the assumption of LTE, we have assumed that the
excitation temperature is equal to the kinetic temperature of 29~K derived from NH$_3$ observations
by Molinari et al.~(\cite{molinari96}), a temperature slightly lower than the dust temperature
derived by fitting the spectral energy distribution (SED) by Chini et al.~(\cite{chini01}) but
similar to the excitation temperature used for other intermediate-mass molecular outflows. We assumed
an [H$_2$]/[CO] abundance ratio of 10$^4$ (e.g., Scoville et al.~\cite{scoville86}), and a $10\%$
Helium abundance. In Table~\ref{tco} we give the mass, the momentum, kinetic energy, and momentum
rate in the outflows, which have been estimated following the derivation of Scoville et
al.~(\cite{scoville86}). We also report the range of outflow velocities for the outflows. Note that
the momentum, kinetic energy, and momentum rate in the outflow have not been corrected for the
(unknown) inclination angle, $i$, of the flow with respect to the plane of the sky. In the case of
correcting for inclination, the velocities should be divided by $\sin i$, and the linear size of the
lobes by $\cos i$. It is almost impossible to separate the emission of both outflows for the SHV
interval, especially close to the center of symmetry of the outflows  (see previous section).
Therefore, it is possible that some emission from outflow B has been counted as part of outflow A and
vice-versa.  As already mentioned in Paper~II, due to some possible filtering of the emission by the
interferometer,  absorption of the emission by the ambient cloud, opacity effects, and the
integration range chosen, the masses calculated should be considered as lower limits. The values of
the mass, momentum, kinetic energy, and momentum rate are consistent with  the values estimated from
interferometric observations for other intermediate-mass molecular outflows (see
Table~\ref{toutflows}). The dynamical timescales of the outflows, which have been estimated
taking the ratio of the maximum separation of the outflow lobes and the terminal velocity of the
outflow measured from the CO line wings, are 3000 and 5000~yr for the blueshifted and redshifted
lobes of outflow A, and 5000 and 7000~yr for the blueshifted and redshifted lobes of outflow B.

%L1641S (see references in Anglada~\cite{anglada95} and Wu
%et al.~\cite{wu04}), or NGC~2071 (Snell et al.~\cite{snell84}), as can be seen in
%Table~\ref{toutflows}, and are more consistent with the values derived for low-mass
%outflows (e.g., Cabrit \& Bertout~\cite{cabrit92}; Anglada~\cite{anglada95}; Bontemps et
%al.~\cite{bontemps96}; Lee et al.~\cite{lee00}, \cite{lee02}). 

\subsection{The dense cores}
\subsubsection{Physical parameters of the dense cores}
\label{gas_prop}

 As can be seen in the spectra (Fig.~\ref{lines}) and the fitted parameters (Table~\ref{table_lines}),
the gas towards OVRO~1 appears to be clearly at a different velocity,  $V_{\rm LSR}$$\simeq$4.7~km\,s$^{-1}$,
than OVRO~2 and OVRO~3, which have velocities of $V_{\rm LSR}$$\simeq$5.5--6\,km\,s$^{-1}$.

Following Eq.~5 of Paper~II, we estimated the virial mass of the sources assuming a spherical
clump with a power-law density distribution $\rho\propto r^{-p}$, with $p=2.0$--1.5, and
neglecting contributions from magnetic field and surface pressure (see Table~\ref{table_lines}).
The virial masses have been computed from C$^{18}$O, except for OVRO~1. For this source, it has
been computed from HC$_3$N as it peaks close to OVRO~1 and is clearly associated with it. For
OVRO~2, the emission is not resolved, and therefore one can only estimate an upper limit. The
values estimated for OVRO~1, 4.3--5.7~$M_{\odot}$, are slightly lower than the mass derived from
the continuum, 6.5~$M_{\odot}$. Taking into account that the mass of the central protostar has not
been considered in this calculation, the total mass (circumstellar plus protostar) of OVRO~1 would
be slightly higher than the virial mass.  This would be consistent with the object undergoing 
collapse, and consistent with accretion driving the outflow. For OVRO~2, from the upper limits computed it is not possible to say whether the
virial mass is higher or lower than the continuum mass. For OVRO~3, the virial mass estimated is
clearly higher than the mass derived from the continuum. However, it is not possible to compare
both masses as the C$^{18}$O emission, from which the virial mass has been computed, is much more
extended than the continuum emission, from which the circumstellar mass has been estimated.

\subsubsection{Envelope mass from gas emission}
\label{gas_emission}

As already mentioned in Paper~II, the gas abundance relative to molecular hydrogen is the main uncertainty
of the estimate of the mass of the molecular core. Therefore, instead of measuring the mass of the gas
towards the protostars from the C$^{18}$O or HC$_3$N emission, assuming a given fractional abundance of
these species, it is better to estimate their fractional abundance from a comparison with the mass derived
from the dust emission. 
We integrated the gas emission in the same  area  used to estimate
the mass of the gas from the continuum dust emission,  and assumed that the mass towards the continuum
sources derived from C$^{18}$O and HC$_3$N is that derived from the 2.7~mm continuum emission.

Following Eq.~3 and 4 of Paper~II, the estimated C$^{18}$O~(\juz) abundance relative to molecular
hydrogen is $2.5\times10^{-8}$ towards OVRO~1 and $\sim$ $5\times10^{-8}$  towards OVRO~2 and OVRO~3.
Typical C$^{18}$O fractional abundances estimated towards molecular clouds are 1.7--4$\times10^{-7}$
(Frerking et al.~\cite{frerking82}; Kulesa et al.~\cite{kulesa05}). Therefore, even taking into account
the uncertainties up to a factor of 5 introduced in the mass estimates by the different dust opacity
laws used, the abundances derived towards OVRO~1, 2 and 3 are lower than the typical fractional
abundances. 

Regarding the HC$_3$N emission, the estimated abundance relative to molecular hydrogen is $1\times10^{-10}$
towards OVRO~1, $6.5\times10^{-11}$  towards OVRO~2, and $8.5\times10^{-11}$  towards OVRO~3.  The 
HC$_3$N  abundance found in the literature can vary by up to 3 orders of magnitude, from 
$\sim$$3\times10^{-11}$ (Vanden Bout et al.~\cite{vandenbout83}) to $\sim$$10^{-8}$ (de Vicente et
al.~\cite{devicente00}; Mart\'{\i}n-Pintado et al.~\cite{martin05}). Therefore, the HC$_3$N abundances derived
towards OVRO~1, and especially those derived towards OVRO~2 and 3 are in the lower end of the range of
fractional abundances estimated toward molecular clouds.

\begin{table*}
\caption[] {Properties of intermediate-mass outflows derived from interferometric observations}
\label{toutflows}
\begin{tabular}{lcccccccc}
\hline
&\multicolumn{1}{c}{$d$}&
\multicolumn{1}{c}{$L$}&
\multicolumn{1}{c}{$M_{\rm env}$}&
\multicolumn{1}{c}{$M_{\rm out}$}&
\multicolumn{1}{c}{$F_{\rm obs}$}&
\multicolumn{1}{c}{$t_{\rm dyn}$}&
\multicolumn{1}{c}{NIR or}
\\
\multicolumn{1}{c}{Outflow}&
\multicolumn{1}{c}{(kpc)}&
\multicolumn{1}{c}{($L_\odot$)}&
\multicolumn{1}{c}{($M_\odot$)}&
\multicolumn{1}{c}{($M_\odot$)}&
\multicolumn{1}{c}{($M_\odot \, \mbox{km s}^{-1}$\,yr$^{-1}$)}&
\multicolumn{1}{c}{($10^3$ yr)}&
\multicolumn{1}{c}{MIR$^{a, b}$}&
\multicolumn{1}{c}{Ref.}
\\
\hline
 IRAS 20343+4129 IRS1 &1.4 &1000--3200 &0.8 &\phantom{1}0.04  &$1.3\times10^{-4}$ 
 &4.2 &Y &1, 2\\
 OMC1-S 137-408$^{c}$ &0.50 & &2 &\phantom{1}0.07  &$1.2\times10^{-2}$  &0.11--3 &N &3, 4, 5\\
 L1206 &0.91 &1200 &14.2 &\phantom{1}0.09  &$7.7\times10^{-5}$ 
 &8.0--20 &Y &6, 7\\
 Cepheus E &0.73 &70 &13.6 &\phantom{1}0.13  &$2.3\times10^{-4}$ &4.0--8.0 &Y  &8, 9, 10 \\
 IRAS 21391+5802 &0.75 &235--500 &5.1 &\phantom{1}0.14  &$1.4\times10^{-3}$  &2.0--3.5 &Y &11, 12,
 13, 14, 15\\
 IRAS 20050+2720 A &0.70 &280 &6.5 &\phantom{1}0.20  &$5.3\times10^{-4}$  &6.5--7.3 &? &16,
 17, 18\\
 OMC-2/3 MM7 &0.45 &76 &0.36--0.72 &\phantom{1}0.34 &$1.1\times10^{-3}$  &6.3--21 &Y &19, 20, 21\\ 
 IRAS 20293+3952 A &2.0 &1050$^{d}$ &4 &2.0  &$2.1\times10^{-2}$  &4.3 &N &1, 22, 23\\
 IRAS 22171+5549 &2.4 &1800 &25 &2.7  &$1.2\times10^{-2}$  &7 &N &24, 25 \\ 
 IRAS 21307+5049 &3.6 &4000 &53  &3.2  &$2.9\times10^{-3}$  &13 &\phantom{.}N$^{e}$ &24, 25 \\
 S235 NNW-SSE &1.8 &&&4.0  &$1.0\times10^{-3}$  &15--20  &N &26, 27 \\
 S235 NE-SW &1.8 &1000   &16 & 9.0 &$1.4\times10^{-2}$  &5.4--14.7 &$?^{f}$ &26, 27 \\
 IRAS 23385+6053 &4.9 &1500 &61 &10.6  &$9.0\times10^{-3}$  &7.3 &N &28, 29\\
 HH~288$^{c}$ &2.0 &500 &6--30 &11 &$1.4\times10^{-2}$  & 28 &N &30 \\
\hline

\end{tabular}

 REFERENCES: 1: Sridharan et al.\  (\cite{srid02}); 2: Palau et al.\ (\cite{palau07b}); 3:  Genzel \&
Stutzki (\cite{genzel89}); 4: Zapata et al.\ (\cite{zapata06}); 5: L.\ A.\ Zapata (private communication);
6: Crampton \& Fisher (\cite{crampton74}); 7: Paper II; 8: Lefloch et al.\ (\cite{lefloch96}); 9:
Moro-Mart\'\i n et al.~(\cite{moro01}); 10: Noriega-Crespo et al. (\cite{noriega04}); 
11: Matthews (\cite{matthews79}); 12: Saraceno et al.\
(\cite{saraceno96}); 13: de Gregorio-Monsalvo (\cite{degregorio06}); 14: Paper I; 15: Getman et al. (\cite{getman07}); 16: 
Dame \& Thaddeus(\cite{dame85}); 17: Froebrich (\cite{froebrich05}); 18: this work; 19: Takahashi et
al.~(\cite{takahashi06});  20: Takahashi et al.~(\cite{takahashi07});  21: Nielbock et al. (\cite{nielbock03});  22: Beuther et al.\
(\cite{beuther04}); 
23: Palau et al.\  (\cite{palau07a}); 24: Molinari et al.\  (\cite{molinari02}); 25:
Fontani et al. (\cite{fontani04a}); 26: Evans \& Blair (\cite{evans81}); 27: Felli et
al.~(\cite{felli04}); 28: Molinari et al.\ (\cite{molinari98}); 29: Fontani et al.\  (\cite{fontani04b});
30: Gueth et al.~(\cite{gueth01}).

(a) Y: source detected at $\lambda<8.0~\mu$m by the
Infrared Array Camera (IRAC) of the {\it Spitzer Space Telescope}, ISOCAM at 7~$\mu$m, or by 2MASS; N:
source not detected at $\lambda<8.0~\mu$m; ?: source for which the association between the millimeter and mid-infrared
emission is dubious. \\
(b) The information was obtained either from the literature of from the Spitzer Science Center Data
Archive interface Leopard. \\
(c) Because of the difficulty of treating each outflow separately, the parameters correspond to the whole
emission, thus including the two outflows associated with OMC1-S 137-408 and with HH288. \\
(d) The luminosity of IRAS~20293+3952 is 6300 $L_\odot$. However, most of its luminosity comes from a
 B1 star ($L$$\sim 5250$  $L_\odot$: Panagia et al.~\cite{panagia73}) associated with it and not from the intermediate-mass YSO powering the
 outflow A (Palau et al.~\cite{palau07a}). \\
(e)  Mid-infrared emission at 7~$\mu$m has been detected towards IRAS~21307+5049 by Fontani et al.~(\cite{fontani04b}). However,
the fact that the mid-infrared emission peak is offset by $\sim$6$''$ from the millimeter and submillimeter
emission peaks led
these authors to suggest that the mid-infrared data are likely due to the nearby cluster. \\
(f) Mid-infrared emission has been observed
$1\farcs5$ to the south of the millimeter peak (Felli et al.~\cite{felli06}). Such a separation is slightly greater than the error of
the relative positions, and according to Felli et al.~(\cite{felli06}) it cannot be firmly established
whether both infrared and millimeter sources are related.\\
 \end{table*}

\begin{figure*}
\centerline{\includegraphics[angle=0,width=18cm]{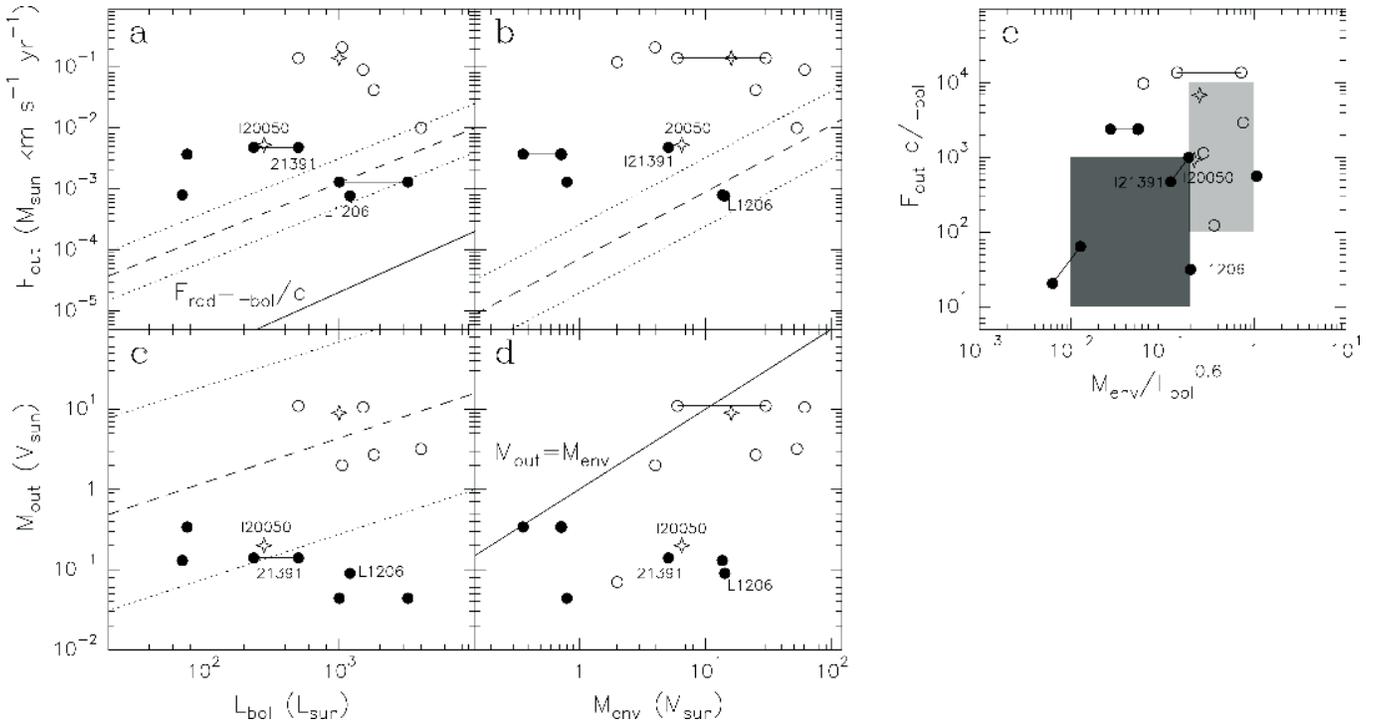}}
\caption{Correlations between source and outflow properties for the sample of intermediate-mass protostars
and outflows  listed in Table~\ref{toutflows}: ({\it a}) Outflow momentum flux $F_{\rm out}$ versus
bolometric luminosity $L_{\rm bol}$. The 'best fit' $F_{\rm out}$--$L_{\rm bol}$ correlation found for
Class~I sources by Bontemps et al.~(\cite{bontemps96}) is plotted as a dashed line. Dotted lines represent
the dispersion of the fit. The solid line  presents the relation $F_{\rm out}=L_{\rm bol}/c$.  ({\it b}) 
$F_{\rm out}$ versus circumstellar envelope mass $M_{\rm env}$. The 'best fit' $F_{\rm out}$--$M_{\rm env}$
correlation found for a low-mass Class~0 and Class~I sample by Bontemps et al.~(\cite{bontemps96}) is
plotted as a dashed line. Dotted lines represent the dispersion of the fit.  ({\it c}) Outflow mass $M_{\rm
out}$ versus $L_{\rm bol}$. The least square linear fit found for a sample of low- to high-mass sources by
Wu et al.~(\cite{wu04}) is plotted as a dashed line. Dotted lines represent the dispersion of the fit. 
({\it d})  ~$M_{\rm out}$ versus $M_{\rm env}$. The solid line 
presents the relation $M_{\rm out}=M_{\rm env}$. ({\it e}) $F_{\rm out}\,c/L_{\rm bol}$ (dimensionless)
versus $M_{\rm env}/L_{\rm bol}^{0.6}$\, ($M_{\rm env}$ and $L_{\rm bol}$ in units of $M_{\odot}$ and
$L_{\odot}$). The light and dark grey boxes roughly denote the zones where the Class~0 and Class~I sources,
respectively, are
located in Fig.~7 of Bontemps et al.~(\cite{bontemps96}).   The horizontal and oblique lines that join two dots
indicate two different values for the same source or outflow property (see Table~\ref{toutflows}).   
Open circles represent sources not detected at $\lambda<8~\mu$m, filled circles represent sources detected
at $\lambda<8~\mu$m, and stars represent sources for which the association between the millimeter and
mid-infrared emission is dubious (see Table.~\ref{toutflows}).  The three sources studied by us in
Paper I, Paper II and this work are labeled.}
\label{out-paper}
\end{figure*}

\section{Intermediate-mass protostars and their outflows}
\label{intermediate-out}

This present work, which is focused on the properties of \iras\ and of its molecular outflow, is
the third of a series of high-angular resolution studies devoted to intermediate-mass protostars and
their outflows, including the study of IRAS~21391+5802 (Paper~I) and IRAS~22272+6358A (Paper~II). We
put the results on \iras\ in the context of our previous results, and in the context of
intermediate-mass star formation. We further compare the properties and evolution of
intermediate-mass protostars with those of their lower mass counterparts. We have compiled
information on outflow properties that  have been derived from interferometric observations in order
to be sensitive to the same spatial scales and to better compare the properties with those derived for the sources in our study.  Our sample is therefore
not very homogeneous,  as very few intermediate-mass outflows have been observed at high-angular
resolution. In any case, we have been able to compile data of a total of 14 intermediate-mass
protostars, including the 3 sources studied by us.

\subsection{Multiplicity in intermediate-mass star-forming regions}

The millimeter continuum emission around \iras\ has been resolved into three millimeter sources (see
Fig.~\ref{cont}): an intermediate-mass source, OVRO~1, and two low-mass sources, OVRO~2 and OVRO~3. In
addition, the powering source of outflow B has not yet been detected at millimeter wavelengths, or its
emission has not been resolved from the extended emission  of the circumstellar envelope surrounding the
intermediate-mass OVRO~1 (see Sect.~\ref{co}). If one also takes into account the millimeter source
detected by Furuya et al.~(\cite{furuya05}), located about $5''$ east of OVRO~1 and not resolved by our
observations, the emission around \iras\ would be associated, at least, with 5 YSOs. And, except for
OVRO~3 that could be associated with \iras\ and with the NIR source~5 detected by Chen et
al.~(\cite{chen97}; see Sect.~\ref{dust}), none of them seems to be detected at infrared wavelengths yet.
This situation very closely resembles  the scenario around the intermediate-mass sources IRAS~21391+5802,
in the bright-rimmed cloud IC~1396N, and IRAS~22272+6358A, in L1206, where the millimeter emission has been
resolved into an intermediate-mass source surrounded by less massive and smaller objects (Paper I; Beltr\'an et
al.~\cite{beltran04a}; Fuente et al.~\cite{fuente07}; Neri et al.~\cite{neri07}; Paper~II).
This suggests that
intermediate-mass stars form in dense clustered environments. In other words, the formation of an
isolated intermediate-mass star if existing appears rare. This is different from the low-mass case, where
examples of protostars formed in relative isolation are known (e.g.\ L1157: Beltr\'an et
al.~\cite{beltran04b}; HH~211: Palau et al.~\cite{palau06a}; or Bok globules: Laundhart et
al.~\cite{laundhart97}). The crowded environments would explain the complexity of the immediate vicinity
of intermediate-mass protostars.

Intermediate-mass protostars are associated with small protoclusters of low-mass millimeter sources,
some of them not yet detected at shorter wavelengths. In many cases, they are also  associated  with young
and embedded clusters of more evolved NIR sources (e.g.\ IRAS~21391+5802: Nisini et al.~\cite{nisini01},
Getman et al.~\cite{getman07}; L1206: Ressler \& Shure~\cite{ressler91},  Kumar et
al.~\cite{kumar06}; IRAS~20050+2720: Chen et a.~\cite{chen97}, Gutermuth et al.~\cite{gutermuth05},
Kumar et al.~\cite{kumar06}; S235: Felli et al.~\cite{felli97}; IRAS~22172+5549:  Fontani et
al.~\cite{fontani04a}, Kumar et al.~\cite{kumar06}; IRAS 21307+5049: Fontani et
al.~\cite{fontani04a};  IRAS 20293+3952 A: Palau et al.~\cite{palau07a}). Interestingly, in 
some cases these NIR  clusters contain a large number of more evolved low-mass Class~II and Class~III
sources (e.g.\ IRAS~21391+5802: Getman et al.~\cite{getman07}, IRAS~20050+2720: Chen et
al.~\cite{chen97};  IRAS 20293+3952 A: Palau et al.~\cite{palau07a}; and possibly L1206: M.\ S.\
N.\ Kumar, private communication). This means that the
intermediate-mass protostars and these more evolved low-mass sources cannot be coeval. All this suggests
 that, at least in some regions, intermediate-mass protostars would start forming after the first
generation of low-mass stars has completed their main accretion phase. Such a situation appears to
be typical of high-mass star-forming regions, where embedded massive stars that are still accreting
material are surrounded by clusters of less embedded and more evolved low-mass Class~I or Class~II sources
(Kumar et al.~\cite{kumar06}).

The presence of multiple sources in intermediate-mass star-forming regions also explains the complex
appearance of the molecular outflows in these regions when observed with low-angular resolution. This is
due to, firstly, the presence usually of more than one outflow in the region (e.g.\ HH~288: Gueth et
al.~\cite{gueth01}; IRAS~21391+5802: Paper~I; \iras: this paper), and secondly, to the stronger interaction
of the outflow with the high-density clumps surrounding the intermediate-mass protostar (IRAS~21391+5802:
Paper~I). However, as shown by these studies of intermediate-mass star-forming regions, the molecular
outflows, which are intrinsically more energetic than those driven by low-mass stars, are collimated and
less complex when observed at high-angular resolution. In fact, they appear collimated even at low outflow
velocities  (Paper~I, Paper~II), similar  to the low-mass protostellar flow in HH~211 (Gueth \&
Guilloteau~\cite{gueth99}; Palau et al.~\cite{palau06b}).

\subsection{Relationships between protostar properties and outflow parameters}
\label{relation}

Table~\ref{toutflows} shows the properties of intermediate-mass outflows and their powering sources
derived from interferometric observations, which have been compiled from the literature. 
Table~\ref{toutflows} also gives information on whether the sources have been detected or not at NIR
or mid-infrared (MIR) wavelengths. As seen in the table, only a very limited number of
intermediate-mass outflows have been studied at high-angular resolution. In spite of the poor
statistics of the sample, we compare the properties of intermediate-mass YSOs with those of low-mass
protostars. To do this, we checked whether intermediate-mass YSOs are consistent with the
correlations between source and outflow properties found for a sample of Class~0 and Class~I
protostars by Bontemps et al.~(\cite{bontemps96}). In particular, we checked the correlation between
circumstellar envelope mass $M_{\rm env}$ and the momentum rate in the outflow $F_{\rm out}$, the
correlation between bolometric luminosity $L_{\rm bol}$ and $F_{\rm out}$, and that between the
normalized momentum $F_{\rm out}\,c/L_{\rm bol}$ and $M_{\rm env}/L_{\rm bol}^{0.6}$\, (see
Fig.~\ref{out-paper}). Following Bontemps et al.~(\cite{bontemps96}), we used the observed momentum
rate in the outflow $F_{\rm obs}$ (see Table~\ref{toutflows}), corrected for inclination and opacity.
We applied a mean correction factor of 2.9 for the inclination, corresponding to a mean inclination
angle with respect to the plane of the sky of $\sim 32.7\degr$, and a mean correction factor of 3.5
for the opacity. Therefore, the corrected momentum flux of the outflow $F_{\rm out}$, is $F_{\rm out}
\sim 2.9 \times 3.5 \times F_{\rm obs} \sim 10\times F_{\rm obs}$.  For the three molecular outflows
that had already been corrected for inclination angle $i$, we only applied the opacity mean
correction factor 3.5. In addition, we also compared $M_{\rm env}$ and the mass of the outflow
$M_{\rm out}$, and $L_{\rm bol}$ and $M_{\rm out}$. The different symbols in Fig.~\ref{out-paper}
indicates  whether the sources have been detected at $\lambda<8~\mu$m or not, or whether the
detection is dubious (see Table~\ref{toutflows}).

 In Fig.~\ref{out-paper}$a$ and \ref{out-paper}$b$, we compare the relationships between $F_{\rm out}$ and
$L_{\rm bol}$ or $M_{\rm env}$\, obtained for the intermediate-mass outflows with those obtained by
Bontemps et al.~(\cite{bontemps96}).  The 'best fit' correlation obtained by these authors is plotted as a
dashed line, and the dispersion of the fit with dotted lines. As can be seen in the plot, intermediate-mass
YSOs have, in general, higher $F_{\rm out}$ than low-mass objects. One could argue that this is due to the
fact that the sources in our sample have been observed with interferometers while those of Bontemps et al.\
with single-dish. However, for those sources in our sample that have been observed with single-dish as
well, the values of $F_{\rm out}$ derived are either comparable or even higher than those derived with
interferometers. According to Eq.~5 of Bontemps et al.~(\cite{bontemps96}), $F_{\rm out}$ is related to the
accretion rate $\dot M_{\rm acc}$, the entrainment efficiency, and the outflow driving engine efficiency.
Therefore, the higher $F_{\rm out}$ values suggest either a higher $\dot M_{\rm acc}$ for intermediate-mass
stars, a higher entrainment efficiency or a higher outflow driving engine efficiency. Calvet et
al.~(\cite{calvet04}) have found, by means of optical and ultraviolet observations, that the average mass
accretion rate for a sample of intermediate-mass T Tauri stars, which are the evolutionary predecessors of
Herbig~Ae stars, is a factor $\sim 5$ higher than that for their low-mass counterparts. Therefore, it is
very likely that intermediate-mass YSOs accrete material faster. This seems a  reasonable conjecture as
these protostars have to gain enough mass before the circumstellar disks are dispersed. According to Fuente
et al.~(\cite{fuente01}),  the dispersal of the circumstellar material and the end of the outflow activity
should occur in $<10^5$~yrs.

As seen in Table~\ref{toutflows} and Fig.~\ref{out-paper},  half of the sources have not been detected at
$\lambda<8.0~\mu$m. This suggests that these sources, represented by open circles in Fig.~\ref{out-paper},
could be more deeply embedded in the core, and therefore to be in an earlier evolutionary stage than those
detected and represented by filled circles.  The difference in detection rate at NIR or MIR wavelengths is
also evident in some source and outflow properties. Fig.~\ref{out-paper} shows that for a similar $L_{\rm
bol}$ or $M_{\rm env}$, the sources not detected at NIR or MIR wavelengths have higher,  in some cases up
to $\sim$2 orders of magnitude higher,  $F_{\rm out}$ and $M_{\rm out}$ than  those detected.  Therefore,
this suggests that the sources not detected at NIR or MIR wavelengths are more efficient at driving their
outflows, which are more powerful, than those detected.   The same happens when comparing $M_{\rm out}$ for
objects with similar $L_{\rm bol}$\, (see Fig.~\ref{out-paper}$c$). Those not detected at NIR or MIR
wavelengths put more mass in the outflows than those detected. In addition, the non-detected sources agree
better with the correlation between $M_{\rm out}$ and $L_{\rm bol}$\, obtained by Wu et al.~(\cite{wu04})
for a sample of low- to high-mass YSOs.  Although the radiation pressure of the central source would not be
enough to drive the outflow if the photons emitted by the central source were scattered only once, as shown
by the fact that all the outflows have $F_{\rm out}$ above the radiative momentum flux $F_{\rm rad}=L_{\rm
bol}/c$\, (see Fig.~\ref{out-paper}$a$), $L_{\rm bol}$\, of the central source and $M_{\rm out}$ are still
correlated, showing the dependence of the outflow on its driving source. As seen in
Fig.~\ref{out-paper}$d$, in general, the sources not detected at NIR or MIR wavelengths put almost the same
amount of mass in the outflow as they have in the circumstellar envelope.

To further investigate these differences, we removed any luminosity dependence from both $F_{\rm out}$ and
$M_{\rm env}$. Following Bontemps et al.~(\cite{bontemps96}), we plotted the outflow efficiency $F_{\rm
out}\,c/L_{\rm bol}$ versus $M_{\rm env}/L_{\rm bol}^{0.6}$\, (Fig.~\ref{out-paper}$e$), along with the
zones where the low-mass Class~0 and Class~I sources are roughly located in Fig.~7 of Bontemps et al.~(\cite{bontemps96}). This plot
shows that, in general, the sources not detected at NIR or MIR wavelengths have a higher outflow efficiency. In the
low-mass regime, there is a decline of the outflow efficiency with evolutionary stage (Bontemps et
al.~\cite{bontemps96}). Therefore, although we do not claim that the sources not detected at $\lambda<8.0~\mu$m are
like low-mass Class 0 objects, we propose that these sources could be younger than those detected, as additionally
suggested by the differences in properties such as $F_{\rm out}$, $M_{\rm out}$, and outflow efficiency. In the
low-mass regime, the youngest protostars have been classified as Class 0 objects, based on criteria such as the
ratio of submillimeter to bolometric luminosity, $L_{\rm submm}/L_{\rm bol}> 5\times10^{-3}$, where $L_{\rm submm}$
is measured longwards of 350~$\mu$m (Andr\'e et al.~\cite{andre93}). However, intermediate-mass star-forming
regions are located, in general, at farther distances than low-mass ones, and usually associated with small
clusters. Hence, the luminosity at wavelengths longwards of 350~$\mu$m will likely have contributions from other
sources in the cluster, and therefore the criterion used to classify the youngest low-mass sources might not be
applicable for intermediate-mass ones. 
In order to confirm this possible evolutionary difference in the sources of our
sample, one should conduct a detailed source by source study of the SED at high-angular resolution, 
however, this goes beyond the scope of this work.

\section{Conclusions}

We studied the dust at 2.7 and 7~mm and the gas at 2.7~mm towards \iras, an intermediate-mass YSO, with
the OVRO Millimeter Array and the VLA. We also put the results on this and previously studied intermediate-mass
sources in the context of intermediate-mass star formation.

The 2.7~mm continuum emission has been resolved into three sources, OVRO~1, OVRO~2, and OVRO~3. Two of them,
OVRO~1 and OVRO~2, have also been detected at 7~mm. OVRO~3, which is located close to the C$^{18}$O emission
peak, could be associated with \iras. The strongest source at millimeter wavelengths is OVRO~1, which has a
deconvolved diameter of $\sim $$3\farcs3$ ($\sim$2300 AU) at the half-peak intensity contour of the dust
emission. The source is surrounded by an extended and a fairly spherical halo or envelope, which has a size of
$\sim$$11\farcs3$ ($\sim$8000 AU) at the 3~$\sigma $ contour level. OVRO~2 has a deconvolved diameter of
$\sim$$2\farcs3$ ($\sim$1600 AU) at the half-peak intensity contour of the dust emission, and OVRO~3 is not
resolved. The mass of the sources, estimated from the dust continuum  emission, is 6.5~$M_{\odot}$ for
OVRO~1, 1.8~$M_{\odot}$ for OVRO~2, and 1.3~$M_{\odot}$ for OVRO~3.  

The multiplicity of sources observed towards \iras, as well as towards IRAS~21391+5802 and L1206,
appears to be typical of intermediate-mass star-forming regions. Intermediate-mass stars form in dense
clustered environments, and in most cases they are associated with small protoclusters of low-mass
millimeter sources, some of them not yet detected at shorter wavelengths. In many cases, as for example
\iras, IRAS~21391+5802 and L1206, intermediate-mass protostars are also  associated  with young and
embedded clusters of more evolved NIR sources that sometimes  contain a large number of more evolved
low-mass Class~II and Class~III sources. This suggests that, at least in some regions, intermediate-mass
protostars would start forming after the first generation of low-mass stars has completed their main
accretion phase, a situation that appears to be typical of high-mass star-forming regions (Kumar et
al.~\cite{kumar06}).

The CO~(\juz) emission towards \iras\ traces two bipolar outflows within the OVRO field of view, a roughly east-west
bipolar outflow, coincident with the outflow A of Bachiller et al.~(\cite{bachiller95}) and apparently
driven by the intermediate-mass source OVRO~1, and a northeast-southwest bipolar outflow, labeled
outflow B and  probably powered by a YSO engulfed in the circumstellar envelope
surrounding OVRO~1. The outflow~A presents a different morphology for the blueshifted and the
redshifted emission. The blueshifted emission is more compact than the redshifted one. This lack of
symmetry between the lobes could be due to the presence of the OVRO~3 core. 
High-angular resolution observations show that, in general, intermediate-mass outflows, which are
intrinsically more energetic than those driven by low-mass YSOs, are not intrinsically more complex than
low-mass outflows. Intermediate-mass outflows appear collimated, even at  low velocities, and have
properties that do not differ significantly from those of low-mass stars.  $F_{\rm out}$ values of
intermediate-mass protostars are higher than those of low-mass ones. Although one cannot discard a higher
entrainment efficiency or outflow driving engine efficiency, this suggests that intermediate-mass YSOs very
likely accrete material faster than low-mass ones.

Intermediate-mass protostars do not form a homogeneous group but have different properties. The sources not
been detected at $\lambda<8.0~\mu$m clearly have higher $F_{\rm out}$, $M_{\rm out}$, and outflow
efficiency than the sources detected at  $\lambda<8.0~\mu$m for similar $M_{\rm env}$ and $L_{\rm bol}$.
This suggests that the sources not detected at NIR or MIR wavelengths, which are likely  more embedded in
the core and drive more powerful outflows, could be in an earlier evolutionary stage.

\begin{acknowledgements}

It is a pleasure to thank the OVRO staff for their support during the observations. MTB is deeply grateful
to Francesco Fontani, Fabrizio Massi, Aina Palau, M.\ S.\ Nanda Kumar, Luis Zapata and Satoko Takahashi for the information provided on
intermediate-mass protostars. MTB, RE, JMG, and GA are
supported by MEC grant AYA2005-08523-C03. JMG acknowledges support from AGAUR grant SGR 00489. GA acknowledges support from Junta de Andaluc\'{\i}a. This
research has made use of the NASA/IPAC Infrared Science Archive, which is operated by the Jet Propulsion
Laboratory, California Institute of Technology, under contract with the National Aeronautics and Space
Administration (NASA). This publication makes use of data products from the Two Micron All Sky Survey, which
is a joint project of the University of Massachusetts and the IPAC/California Institute of Technology,
funded by  NASA and the National Science Foundation.

 \end{acknowledgements}


\begin{thebibliography}{}	        		      
  
\bibitem[1993]{andre93}
 Andr\'e, P., Ward-Thompson, D., \& Barsony, M.\ 1993, ApJ, 406, 122

 
\bibitem[1998a]{anglada98a} 
 Anglada, G., Rodr\'\i guez, L.\ F., \& Torrelles, J.\ M.\ 1998a, in ASP Conf.\ Ser.\ 132,
 Star Formation with the Infrared Space Observatory, ed.\ J.\ L.\ Yun \& R.\ Liseau (San
 Francisco: ASP), 303
 

\bibitem[2001]{arce01} 
 Arce, H.\ G., \& Goodman, A.\ A.\ 2001, ApJ, 554, 132 

\bibitem[1996]{bachiller96} 
 Bachiller, R.\ 1996, ARA\&A, 34, 111

\bibitem[1995]{bachiller95} 
 Bachiller, R., Fuente, A., \& Tafalla, M.\ 1995, ApJ, 445, L51 


\bibitem[2006a]{beltran06a}				      
 Beltr\'an, M.\ T., Brand, J., Cesaroni, R., Fontani, F., Pezzuto, S., Testi, L., Molinari, S.\ 2006a, A\&A, 447,
 221 	
 

\bibitem[2006b]{beltran06b}				      
 Beltr\'an, M.\ T., Girart, J.\ M., \& Estalella, R.\ 2006b, A\&A, 457, 865 (Paper II)							   
	 
\bibitem[2002]{beltran02}				      
 Beltr\'an, M.\ T., Girart, J.\ M., Estalella, R., Ho, P.\ T.\ P., \& Palau, A.\ 2002, ApJ, 573, 246 (Paper I)							   

\bibitem[2004a]{beltran04a}				       
 Beltr\'an, M.\ T., Girart, J.\ M., Estalella, R., \& Ho, P.\ T.\ P.\ 2004, A\&A, 426, 941							    

\bibitem[2004b]{beltran04b}				       
 Beltr\'an, M.\ T., Gueth, F., Guilloteau, S., \& Dutrey, A.\ 2004, A\&A, 416, 631

\bibitem[2004a]{beuther04}				       
Beuther, H., Schilke, P., \& Gueth, F.\ 2004, ApJ, 608, 330

\bibitem[1996]{bontemps96}
Bontemps, S., Andr\'e, P., Terebey, S, \& Cabrit, S.\ 1996, A\&A, 311, 858

\bibitem[1995]{briggs95}
 Briggs, D.\ 1995, Ph.D.\ Thesis, New Mexico Inst.\ Mining \& Tech. 


\bibitem[2004]{calvet04}
Calvet, N., Muzerolle, J., Brice\~no, C., Hern\'andez, J.\ et al.\ 2004, ApJ, 128, 1294  

\bibitem[1997]{chen97}
 Chen, H., Tafalla, M., Greene, T.\ P., Myers, P.\ C., \& Wilner, D.\ J.\ 1997, ApJ, 475, 163 


\bibitem[2001]{chini01}
 Chini, R., Ward-Thompson, D., Kirk, J.\ M., Nielbock, M., Reipurth, B., \& Sievers, A.\ 2001, A\&A, 369, 155

\bibitem[1999]{choi99}
Choi, M., Panis, J.-F., \& Evans II, N.\ J., 1999, A\&ASS, 122, 519

\bibitem[1974]{crampton74} 
Crampton, D., \&  Fisher, W.\ A.\ 1974, Publ.\ Dominion Ap.\ Obs.\ Victoria, 14


\bibitem[1985]{dame85}
Dame, T.\ M., \& Thaddeus, P.\ 1985, ApJ, 297, 751

\bibitem[2006]{degregorio06}
de Gregorio-Monsalvo, I., G\'omez, J.\ F., Su\'arez, O., Kuiper, T.\ B.\ H., Rodr\'{\i}guez, L.\
F., \& Jim\'enez-Bail\'on, E.\ 2006, ApJ, 642, 319

\bibitem[2000]{devicente00} 
de Vicente, P., Mart\'{\i}n-Pintado, J., Neri, R., \& Colom, P.\ 2000, A\&A, 361, 1058

\bibitem[1981]{evans81}
Evans; N.\ J.\ II, \& Blair, G.\ N.\ 1981, ApJ, 246, 394 

\bibitem[1982]{frerking82}
Frerking, M.\ A., Langer, W.\ D., \& Wilson, R.\ W.\ 1982, ApJ, 262, 590

\bibitem[2004]{felli04}
 Felli, M., Massi, F., Navarrini, A., Neri, R., Cesaroni, R., \& Jenness, T.\
 2004, A\&A, 420, 553

\bibitem[2006]{felli06}
 Felli, M., Massi, F., Robberto, M., \& Cesaroni, R.\ 2006, A\&A, 453, 911

\bibitem[1997]{felli97}
 Felli, M., Testi, L., Valdettaro, R., \& Wang, J.-J.\ 1997, A\&A, 320, 594
 2002, A\&A, 420, 553

\bibitem[2004a]{fontani04a}
Fontani, F., Cesaroni, R., Testi, L., Molinari, S., Zhang, Q., Brand, J., \& Walmsley, C.\ M.\
2004a, A\&A, 424, 179 

\bibitem[2004b]{fontani04b}
Fontani, F., Cesaroni, R., Testi, L., Walmsley, C.\ M.\ et al.\ 2004b, A\&A, 414, 299

\bibitem[2005]{froebrich05}
 Froebrich, D.\ 2005, ApJSS, 156, 169

\bibitem[2007]{fuente07}
 Fuente, A., Ceccarelli, C., Neri, R., Alonso-Albi, T.\ et al.\ 2007, A\&A, 468, L37 

\bibitem[2001]{fuente01}
 Fuente, A., Neri, R., Mart\'{\i}n-Pintado, J., Bachiller, R., Rodr\'{\i}guez-Franco, A., \& Palla, F.\
 2001, A\&A, 366, 873

\bibitem[2003]{furuya03}
 Furuya, R.\ S., Kitamura, Y., Wootten, A., Claussen, M.\ J., Kawabe, R.\ 2003, ApJSS, 144, 71

\bibitem[2005]{furuya05}
 Furuya, R.\ S., Kitamura, Y., Wootten, A., Claussen, M.\ J., Kawabe, R.\ 2005, A\&A, 438, 571


\bibitem[1989]{genzel89}
Genzel, R., \& Stutzki, J.\ 1989, ARA\&A, 27, 41 

\bibitem[2007]{getman07}
Getman, K., Feigelson, E.\ D., Garmire, G., Broos, P., \&  Wang, J.\ 2007, ApJ, 654, 316

\bibitem[1999]{gueth99}
 Gueth, F., \& Guilloteau, S.\ 1999, A\&A, 343, 571
 
\bibitem[1997]{gregersen97}
 Gregersen, E.\ M., Evans, N.\ J.\ II, Zhou, S., \& Choi, M.\ 1997, ApJ, 484, 256
 
\bibitem[2001]{gueth01}
 Gueth, F., Schilke, P., \& McCaughrean, M.\ J.\ 2001, A\&A, 375, 1018

\bibitem[2005]{gutermuth05}
 Gutermuth, R.\ A., Megeath, T., Pipher, J.\ L., Williams, J.\ P.\ et al.\ 2005, ApJ, 632, 397 
 

\bibitem[1983]{hildebrand83}
Hildebrand, R.\ H.\ 1983, QJRAS, 24, 267

\bibitem[1997]{hogerheijde97}
 Hogerheijde, M.\ R., van Dishoeck, E.\ F., Blake, G.\ A., \& van Langevelde, H.\ J.\ 1997, ApJ, 489, 293

\bibitem[1999]{hogerheijde99}
 Hogerheijde, M.\ R., van Dishoeck, E.\ F., Salverda, J.\ M., \& Blake, G.\ A.\ 1999, ApJ, 513, 350


\bibitem[2005]{kulesa05}
Kulesa, C.\ A., Hungerford, A.\ L., Walker, C.\ K., Zhang, X., \& Lane, A.\ P.\ 2005, ApJ, 625, 194

\bibitem[2006]{kumar06}
Kumar, M.\ S.\ N., Keto, E., \& Clerkin, E.\ 2006, A\&A, 449, 1033

\bibitem[1997]{laundhart97}
Laundhart,, R., \& Henning, Th.~ 1997, A\&A, 326, 329 

\bibitem[1996]{lefloch96}
Lefloch, B., Eisl\"offel, J., \& Lazareff, B.\ 1996, A\&A, 313, L17

\bibitem[2000]{looney00}
 Looney, L.\ W., Mundy, L.\ G., \& Welch, W.\ J.\ 2000, ApJ, 529, 477


\bibitem[1998]{macleod98}
MacLeod, G.\ C., Scalise, E.\ Jr., Saedt, S., Galt, J.\ A., \& Gaylard, M.\ J.\ 1998, AJ, 116, 1897


\bibitem[2005]{martin05}
Mart\'\i n-Pintado, J., Jim\'enez-Serra, I., Rodr\'\i guez-Franco, A., Mart\'\i n, S., \&
Thum, C.\ 2005, ApJ, 628, L61

\bibitem[1979]{matthews79}
Matthews, T.\ J.\ 1979, A\&A, 75, 345


\bibitem[1996]{molinari96}
Molinari, S., Brand, J., Cesaroni, R., \& Palla, F.\ 1996, A\&A, 308, 573

\bibitem[1998]{molinari98}
Molinari, S., Testi, L., Brand, J., Cesaroni, R., \& Palla, F.\ 1998, ApJ, 505, L39

\bibitem[2002]{molinari02}
Molinari, S., Testi, L., Rodr\'\i guez, L.\ F., \& Zhang, Q.\ 2002, ApJ, 570, 758

\bibitem[2001]{moro01}
Moro-Mart\'\i n, A., Noriega-Crespo, A., Molinari, S., Testi, L., Cernicharo,
J., \& Sargent, A.\ I.\  2001, ApJ, 555, 146

\bibitem[2007]{neri07}
Neri, R., Fuente, A., Ceccarelli, C., Caselli, P. et al.\ 2007, A\&A, 468, L33 	

\bibitem[2003]{nielbock03}
Nielbock, M., Chini, R., \& M\"uller, S.\ A.\ H.\ 2003, A\&A, 408, 245

\bibitem[2001]{nisini01}
Nisini, B., Massi, F., Vitali, Giannini, T.\ et al.\ 2001, A\&A, 376, 553

\bibitem[2004]{noriega04}
Noriega-Crespo, A., Moro-Mart\'{\i}n, A., Carey, S., Morris, P.\ W.\ et al.~2004, ApJSS, 154, 402


\bibitem[1994]{ossenkopf94}
 Ossenkopf, V., \& Henning, Th.\ 1994, A\&A, 291, 943

\bibitem[2006a]{palau06a}
Palau, A.~2006, Ph.D.\ Thesis, Universitat de Barcelona

\bibitem[2007a]{palau07a}
Palau, A., Estalella, R., Girart, J.\ M., Ho, P.\ T.\ P., Zhang, Q., \& Beuther, H.
2007a, A\&A, 465, 219

\bibitem[2007b]{palau07b}
Palau, A., Estalella, R., Ho, P.\ T.\ P., Estalella, R.\ 2006b, A\&A, 636, L137

\bibitem[2006b]{palau06b}
Palau, A., Ho, P.\ T.\ P., Zhang, Q., Beuther, H., \& Beltr\'an, M.\ T.\
2007b, A\&A, submitted

\bibitem[1991]{palla91}
Palla, F., Brand, J., Cesaroni, R., Comoretto, G., \& Felli, M.\ 1991, A\&A, 246, 249 

\bibitem[1973]{panagia73}
Panagia, N.\ 1973, AJ, 78, 929


\bibitem[1991]{ressler91}
 Ressler, M.\ E., \& Shure, M.\ 1991, AJ, 102, 1398


\bibitem[2000]{richer00}
 Richer, J.,  Shepherd, D.\ S., Cabrit, S., Bachiller, R., \& Churchwell, E.\ 2000, in Protostars and
Planets IV, ed. V.\ Mannings, A.\ Boss, \& S.\ Russell (Tucson: University of Arizona Press), 867


\bibitem[2003]{ridge03}
 Ridge, N.\ A., Wilson, T.\ L., Megeath, S.\ T., Allen, L.\ E., \& Myers, P.\ C.\ 2003, AJ, 126, 286

\bibitem[1996]{saraceno96}
 Saraceno, P., et al.\  1996, A\&A, 315, L293

\bibitem[1993]{scoville93} 
Scoville, N.\ Z., Carlstrom, J.\ E., Chandler, C.\ J., Phillips, J.\ A.\ et al.\ 1993, PASP, 105, 1482

\bibitem[1986]{scoville86} 
 Scoville, N.\ Z., Sargent, A.\ I., Sanders, D.\ B., Claussen,
 M.\ J., Masson, C.\ R., Lo, K.\ Y., \& Phillips, T.\ G.\ 1986, ApJ, 303, 416 

\bibitem[2005]{shepherd05}
 Shepherd, D.\ S.\ 2005, in Massive  Star Birth: A Crossroads of Astrophysics, ed.\ R.\ Cesaroni,
 M.\ Felli, E.\ Churchwell, \& M.\ Walmsley (Cambridge: Cambridge University Press), Proc.\ IAU Symp.\ 227, 237   

\bibitem[1996a]{shepherd96a}
 Shepherd, D.\ S., \& Churchwell, E.\ 1996a, 457, 267

\bibitem[1996b]{shepherd96b}
 Shepherd, D.\ S., \& Churchwell, E.\ 1996b, 472, 225

\bibitem[2006]{schneider06}
Schneider, N., Bontemps, S., Simon, R., Jakob, H., Motte, F.\ et al.\ 2006, A\&A, 458, 855

\bibitem[2002]{srid02}
Sridharan, T.\ K., Beuther, H., Schilke, P., Menten, K.\ M.., \& Wyrowski, F.\ 2002, ApJ, 566, 931


\bibitem[2006]{takahashi06}
Takahashi, S., Saito, M., Takakuwa, S., \& Kawabe, R.\ 2006, ApJ, 651, 933

\bibitem[2007]{takahashi07}
Takahashi et al.\ 2007, in preparation

\bibitem[1983]{vandenbout83} 
Vanden Bout, P.\ A., Loren, R.\ B., Snell, R.\ L., \& Wootten, A.\ 1983, ApJ, 271, 161

\bibitem[1989]{wilking89}
Wilking, B.\ A., Mundy, L.\ G., Blackwell, J.\ H., \& Howe, J.\ E.\ 1989, ApJ, 345, 257 

\bibitem[1999]{williams99}
Williams, J.\ P., \& Myers, P.\ C.\ 1999, ApJ, 511, 208 


\bibitem[2004]{wu04}
Wu, Y., Wei, Y., Zhao, M., Shi, Y., Yu, W.\ et al.\ 2004,  A\&A, 426, 503

\bibitem[2006]{zapata06}
Zapata, L.\ A., Ho, P.\ T.\ P., Rodr\'\i guez, L.\ F., O'Dell, C.\ R., Zhang, Q., \& Muench, A.\
2006, ApJ, 653, 398

\bibitem[2001]{zhang01}
Zhang, Q., Hunter, T.\ R., Brand, J., Sridharan, T.\ K.\ et al.\ 2001, ApJ, 552, 167

\bibitem[2005]{zhang05}
Zhang, Q., Hunter, T.\ R., Brand, J., Sridharan, T.\ K.\ et al.\ 2005, ApJ, 625, 864


\end{thebibliography}
\end{document}